%% file: P02a.tex
\documentclass[dvips]{aa}
\usepackage{txfonts}
\usepackage{graphicx}
\usepackage[colorlinks,citecolor=blue]{hyperref}

\def\logg{\log g}
\def\logL{\log L}
  \def\Teff{T_{\rm\kern-.15em ef\kern-.05em f}} 
\def\vinf{v_\infty}
\def\Mdot{\dot M}
\def\Msun{M_\odot}
\def\Rsun{R_\odot}
\def\Lsun{L_\odot}

\graphicspath{{plots/}}

\begin{document}

\title{Radiation-driven winds of hot luminous stars}
\subtitle{XV.
Constraints on the mass--luminosity relation\\
of central stars of planetary nebulae}

\author{A.\,W.\,A.\,Pauldrach\inst{1} \and
T.\,L.\,Hoffmann\inst{1} \and
R.\,H.\,M\'endez\inst{2}}

\institute{Institut f\"{u}r Astronomie und Astrophysik
der Universit\"{a}t M\"{u}nchen,
Scheinerstra{\ss}e~1, 81679~M\"{u}nchen, Germany \and
Institute for Astronomy, University of Hawaii,
2680~Woodlawn Drive, Honolulu, HI~96822, U.S.A.}

\offprints{A.\,W.\,A.\,Pauldrach,\\
http://www.usm.uni-muenchen.de/people/adi/adi.html}

\date{Received date / Accepted date}

\abstract{ We present a new model atmosphere analysis of nine central
stars of planetary nebulae. This study is based on a new generation of
realistic stellar model atmospheres for hot stars; state-of-the-art,
hydrodynamically consistent, spherically symmetric model atmospheres
that have been shown to correctly reproduce the observed UV spectra
of massive Population~I O-type stars. The information provided by the
wind features (terminal velocity, mass loss rate) permits to derive
the physical size of each central star, from which we can derive the
stellar luminosity, mass, and distance, without having to assume a
relation between stellar mass and luminosity taken from the theory
of stellar structure and AGB and post-AGB evolution. The results of
our analysis are quite surprising: we find severe departures from
the generally accepted relation between post-AGB central star mass
and luminosity.

}

\titlerunning{Radiation-driven winds of PN central stars}
\authorrunning{Pauldrach et al.}
\maketitle

\section{Introduction}
\label{sec:introduction}

In recent years there has been substantial progress in the modelling
of expanding atmospheres of hot stars. It is now possible to produce
synthetic UV spectra of O~stars that resemble the real, observed
ones nearly perfectly. The state-of-the-art wind models deal with
radiatively driven, homogeneous, stationary, extended, outflowing,
spherically symmetric atmospheres. A complete model atmosphere
calculation involves solving the hydrodynamics and the NLTE problem
(rate equations for all important elements, radiative transfer,
and energy equation). The solution of the total interdependent
system of equations has thereby to be based on a non-restrictive
treatment. This permits the calculation of the predicted or
synthetic spectrum, which is then compared to the observed UV
spectrum (cf.~Pauldrach et al., \cite{Pauldrach-et-al-2001}). The
process is repeated adopting different stellar parameters until a
satisfactory fit is obtained. In this kind of work it is not necessary
to adopt an arbitrary velocity law for the wind; the solution is
hydrodynamically consistent and gives us the velocity law as well as
the mass-loss rate (cf.~Pauldrach~\cite{Pauldrach-2003}, Pauldrach
and Hoffmann~\cite{Pauldrach-Hoffmann-2003}).

A very important consequence of these recent developments is that the
fits to the UV spectral features provide information about all the
basic stellar parameters: effective temperature ($\Teff$), radius ($R$)
-- or equivalently, luminosity ($L$) --, mass of the star ($M$), terminal
wind velocity ($\vinf$), and mass loss rate ($\Mdot$). Therefore we
have a purely spectroscopic method to obtain separately $L$ and $M$.

Using this new generation of realistic stellar model atmospheres, Pauldrach et
al.~(\cite{Pauldrach-et-al-2001}), Pauldrach~(\cite{Pauldrach-2003}), and
Pauldrach and Hoffmann~(\cite{Pauldrach-Hoffmann-2003}) present analyses of the
massive O~supergiants HD~30614 ($\alpha$~Cam) and HD~66811 ($\zeta$~Pup) which
provide excellent matches to the observable UV-spectra, thus determining the
basic stellar parameter sets of these objects. Further examples of this kind of
work are given by Hoffmann and Pauldrach~(\cite{Hoffmann-Pauldrach-2001}), who
confirm in their analysis of a subsample of galactic massive O~stars the
parameters derived by Puls et al.~(\cite{Puls-et-al-1996}) from an optical
investigation.

Since this method produces reasonable results when applied to massive
Population~I stars, we now want to apply it to another kind of hot stars: the
central stars of planetary nebulae (CSPNs in what follows). This permits, for the
first time, to test the predictions from post-AGB evolutionary calculations. (The
idea is described together with a first application by Pauldrach et
al.~\cite{Pauldrach-et-al-1988}; preliminary results of the present investigation
have be published by Pauldrach et al.~\cite{Pauldrach-et-al-2001a}.) The earlier
work on CSPNs, based on plane-parallel non-LTE model atmospheres (e.g., M\'endez
et al.~\cite{Mendez-et-al-1988a}, \cite{Mendez-et-al-1988b}) could not provide a
completely independent test, in the following sense: since the plane-parallel
model fits to H and He photospheric absorption lines can only produce information
about surface temperature, He abundance and surface gravity ($\logg$), we cannot
derive stellar masses or luminosities, but only $L/M$ ratios. This is exactly the
same problem we face when dealing with low-gravity early-type ``supergiant''
stars at high Galactic latitudes: are they luminous and massive, or are they
evolving away from the AGB? We need some independent evidence to settle the issue
-- for example, the distance to the star. Unfortunately, we lack reliable
distances to almost all CSPNs.

What could be done was to plot the positions of CSPNs in
the $\logg$--$\log\Teff$ diagram, and compare them with plots of
post-AGB tracks, translated from the $\logL$--$\log\Teff$ diagram.
After doing this translation it is possible to read the stellar mass
in the $\logg$--$\log\Teff$ diagram. From this, we can derive $L$
and, if we know the visual dereddened apparent magnitude, a so-called
``spectroscopic distance''. All this work, however, is based on {\em
assuming\/} that the evolutionary models give us the correct relation
between stellar mass and luminosity. It is not a real test of the
evolutionary models, but only a consistency check.

The new models allow us to overcome this limitation and produce for the first
time information on $L$ and $M$ which is completely independent from the theory
of stellar structure and evolution. In this paper we present the initial outcome
of the project. We have made a careful selection of CSPNs for which we have
adequate spectrograms covering both the visible and UV spectrum. In
Sections~\ref{sec:uv-spectra} and~\ref{sec:optical-spectra} we describe the
available spectrograms. In Section~\ref{sec:procedure} we present the necessary
information about the new wind models and we outline the fitting procedure.
Section~\ref{sec:theory} introduces the relation between wind momentum loss rate
and stellar luminosity, predicted by the theory of radiatively driven winds, and
briefly describes earlier efforts to verify if the CSPNs follow this relation.
Then in Sections~\ref{sec:cspn-detail} and~\ref{sec:cspn-rest} we present the UV
spectral fits using the new wind models, explaining what discrepancies there are
with respect to the earlier modelling and producing a table with the stellar
parameters determined. Section~\ref{sec:interpretation} gives the interpretation
of the CSPN winds and a discussion of the results concerning stellar luminosities
and masses. In Section~\ref{sec:other-observations} we estimate the spectroscopic
distances and compare them with other distance determinations, with inconclusive
results. Section~\ref{sec:masses} deals with other estimates of pre-white dwarf
masses. In Section~\ref{sec:conclusions} we summarize the conclusions.

\section{The UV spectrograms}
\label{sec:uv-spectra}

The sample of CSPNs we have analyzed is defined by the availability
of adequate high-resolution UV spectra, covering the spectral region
between 1000 and 2000~\AA.

With the exception of that of He~2-108, all UV~spectra of our
sample were obtained from the INES Archive Data Server on the Web at
\href{http://ines.laeff.esa.es/}{http://ines.laeff.esa.es/}, providing
access to IUE Final Archive data.  Apart from rectification ``by eye''
(aided by our experience with UV~spectra from massive O~stars),
no further processing was done on the spectra.

The spectrum of He~2-108 is an HST/FOS spectrum (Proposal ID~5339,
PI~RHM) rectified by Haser~(\cite{Haser-1995}). Here, interstellar
Lyman-$\alpha$ absorption has been taken into account in the
rectification process, leading to an empty band around 1216~\AA\
with noisy edges.

\section{The optical spectrograms}
\label{sec:optical-spectra}

Most of the optical spectrograms were obtained by one of us (RHM)
during two observing runs at the European Southern Observatory (ESO),
Cerro La Silla, on 21--24 March and 13--17 June~1994 (the first one
by remote control from ESO Garching), using the 3.5~m New Technology
Telescope (NTT) and EMMI spectrograph in dichroic mode. In this mode
of operation a dichroic beam-splitter was inserted, allowing to obtain
blue and red spectrograms simultaneously.

A long slit of 1~arc~sec width was used in the focal plane.
The slit was oriented either N-S or E-W.

The dispersing element in the blue arm was grating number~3, with
1200~lines/mm, and the camera was $f/4$, giving a pixel size of
0.37~arc~sec and a dispersion of 0.42~\AA/pixel. The detector was
a Tektronix CCD, $1124 \times 1024$ pixels, with pixel size of
24~$\mu$m. In order to decrease the readout time only the central
200~pixels along the slit were read. Therefore the resulting
spectrograms covered 74~arc~secs along the slit, and 450~\AA\ in
wavelength, with central wavelengths of 4500~\AA\ and 4535~\AA\
in March and June~1994, respectively.

The dispersing element in the red arm was grating number~6, with
1200~lines/mm, and the camera was $f/5.2$, giving a pixel size
of 0.27~arc~sec and a dispersion of 0.31~\AA/pixel. The detector
was a Tektronix CCD, $2086 \times 2048$ pixels, with pixel size of
24~$\mu$m. In order to decrease the readout time only the central
400~pixels along the slit were read. Therefore the resulting
spectrograms covered 108~arc~secs along the slit, and 650~\AA\
in wavelength, with central wavelengths of 6560~\AA\ and 6505~\AA\
in March and June~1994, respectively.

The exposure times were short in the red arm, to avoid
saturation of the strongest nebular emission lines.

The usual calibration frames (bias, dome flats, He-Ar comparison
spectrum for wavelength calibration) were obtained for the observations
in both arms. The CCD reductions were made using standard tasks
provided in IRAF\footnote{IRAF is distributed by the National
Optical Astronomical Observatories, operated by the Association of
Universities for Research in Astronomy, Inc., under contract to
the National Science Foundation of the U.S.A.}. After bias-level
subtraction and flatfielding, whenever possible the spectrograms were
registered and combined, to eliminate cosmic ray events and improve
the signal-to-noise ratio. The image combination was made with the
IRAF task {\em imcombine}, using the {\em average} option and {\em
ccdclip} rejection (pixel rejection based on CCD noise parameters).


\section{The modelling procedure}
\label{sec:procedure}

The UV spectrum between 1000 and 2000~\AA\ carries a lot of
information: P-Cygni-type profiles of resonance lines of several ions
of C, N, O, Si, S, P, as well as hundreds of strongly wind-contaminated
lines of Fe\,{\sc iv}, Fe\,{\sc v}, Fe\,{\sc vi}, Cr\,{\sc v},
Ni\,{\sc iv}, Ar\,{\sc v}, Ar\,{\sc vi}. But the information about
the stellar parameters can be extracted only after careful analysis.
A very important recent improvement of our method concerns the
development of a substantially consistent treatment of the blocking
and blanketing influence of all metal lines in the entire sub- and
supersonically expanding atmosphere. All the results we will present
are based on this new generation of models.

The analysis method is based on modelling a homogeneous, stationary,
extended, outflowing, spherically symmetric radiation-driven
atmosphere. A detailed description is given by Pauldrach et
al.~(\cite{Pauldrach-et-al-2001}), Pauldrach~(\cite{Pauldrach-2003}),
and Pauldrach and Hoffmann~(\cite{Pauldrach-Hoffmann-2003}).

The procedure is as follows. A preliminary inspection of the visual
and/or UV spectrum of the star to be analyzed gives an estimate
of $\Teff$. From the UV spectrum, the terminal wind velocity $\vinf$
can be measured directly. Now initial values for the stellar radius $R$
(defined at a Rosseland optical depth of $2/3$) and for the stellar
mass $M$ are assumed.

With the current values of $R$, $\Teff$, $M$, and assuming a set
of abundances, we can solve the model atmosphere and calculate the
velocity field, the mass loss rate $\Mdot$, and the synthetic spectrum.
If the calculated terminal wind velocity $\vinf$ of the model differs
from the observed one, we modify $M$ until agreement is reached
(since $\vinf$ scales with $(M/R)^{1/2}$ according to the theory of
radiation-driven winds). Now the predicted spectrum is compared to
the observed one. If the fit is not satisfactory, we need to modify
$\Mdot$ via a change of $R$ (since $\log \Mdot \sim \log L$, according
to radiation-driven wind theory). The change in $R$ forces us to
change the mass, too, in order to keep $\vinf$ consistent with the
observed value. The new model is calculated and the process is repeated
until we obtain a good fit to all features in the observed spectrum.
(Additionally, $\Teff$ might need to be corrected slightly during
this iteration, if the initial guess was not satisfactory.)

With this procedure our current models produce satisfactory results
for massive Population~I stars. What happens if we apply the same
procedure to CSPNs?

\section{The wind properties of hot stars}
\label{sec:theory}

As a first point of our investigation we examine the dynamical parameters $\vinf$
and $\Mdot$ of radiation-driven CSPN winds.

The significance of this parameters is obvious, since it is the consistent
hydrodynamics which provides the link between the stellar parameters ($\Teff$,
$M$, $R$) and the appearance of the UV spectra, because the latter are determined
by the interplay of the NLTE model and the hydrodynamics. The link is the line
force which controls the hydrodynamics, and which is controlled by the occupation
numbers, and the radiative transfer of the NLTE model. The hydrodynamics in turn
affects the NLTE model and thus the spectra via the density and velocity
structure.

\subsection{The relation between wind-momentum loss rate and luminosity}

A tool for illustrating the behavior of the dynamical parameters is
offered by the so called {\it wind-momentum--luminosity relation}.

\begin{figure}
\includegraphics[angle=-90,width=\columnwidth]{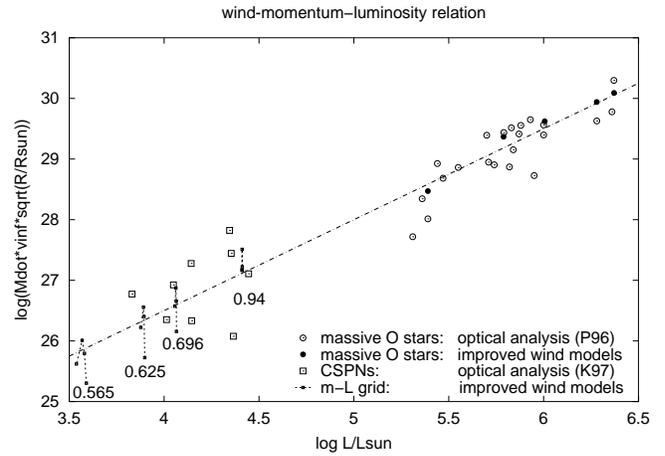}
\caption{
The wind-momentum--luminosity relation for massive O~stars
and CSPNs. P96 designates the analysis based on
H$\alpha$ profiles by Puls et al.~\cite{Puls-et-al-1996},
K97 that of CSPNs by Kudritzki et al.~\cite{Kudritzki-et-al-1997}.
Also plotted are the calculated wind momenta for a sample of
massive O~stars and for a grid of stars following post-AGB
evolutionary tracks (masses given in $\Msun$).}
\label{fig:wml1}
\end{figure}

The significance of this relation is based on the fact that, due to the
driving mechanism of hot stars, the mechanical momentum of the wind
flow ($\vinf \Mdot$) is mostly a function of photon momentum ($L/c$)
and is therefore related to the luminosity.  Thus, the radiatively
driven wind theory predicts, for fixed abundances, a simple relation
between the quantity $\Mdot \vinf$, which has the dimensions of a
momentum loss rate, and the stellar luminosity:
$$
\Mdot \vinf \sim R^{-1/2} L^{1/\alpha}
$$
where $\alpha$, related to the power law exponent of the line
strength distribution function, is \mbox{$\simeq 2/3$} (slightly
dependent on temperature and metallicity; see, for example, Puls et
al.~\cite{Puls-et-al-1996}).  As the expression $\vinf \Mdot R^{1/2} $
is an almost directly observable quantity (see below), it is practical
to plot the log of $\Mdot \vinf R^{1/2}$ as a function of $\log L$.
In this kind of plot the theory predicts, in first approximation,
a linear relation, which is indeed followed by all kinds of massive
hot stars, as shown in Figure~\ref{fig:wml1}.


An initial attempt to verify if CSPNs follow the
wind-momentum--luminosity relation was partly successful (see
Figure~3 in Kudritzki et al.~\cite{Kudritzki-et-al-1997} and also our
Figure~\ref{fig:wml1}). In that paper, terminal wind velocities $\vinf$
were taken from observed UV~spectra and $Q$-values (a quantity relating
mass loss rate and stellar radius, $Q \sim \Mdot (R \vinf)^{-3/2}$)
were derived from observed H$\alpha$ profiles\footnote{We
refer to this as ``optical analysis'', since although
$\vinf$ was taken from UV~spectra, this is a quantity that can be
derived easily and does not require much analysis. The real analysis
determining $Q$ and $\logg$ using model atmospheres was performed
using optical spectra. } using the optical spectra described in
section~\ref{sec:optical-spectra}. Stellar masses were derived
from $\Teff$ and $\logg$, using post-AGB tracks plotted in the
$\logg$--$\log \Teff$ diagram. The stellar radii (and thus, mass loss
rates) and luminosities were then obtained from the masses and the
post-AGB mass--luminosity relation. The CSPNs were found to be at
the expected position along the wind-momentum--luminosity relation,
indicating a qualitatively successful prediction by the theory of
radiatively driven winds. However, the situation was not satisfactory
because there appeared to be a large dispersion in wind strengths
at a given luminosity (strong-winded and weak-winded CSPNs) and some
of the CSPN masses and luminosities were very high ($M>0.8\,\Msun$),
in contradiction with theoretical post-AGB evolutionary speeds.

Thus, at that point we had a qualitatively positive result, namely
that in principle the CSPN winds appear to obey the same physics as
the massive O~star winds; but we also had some unsolved problems.

This situation has been recently discussed by Tinkler and
Lamers~(\cite{Tinkler-Lamers-2002}), who try to improve the central
star parameters by imposing consistency between the evolutionary age of
the central star and the dynamical age of its PN. As result of scaling
the distances and stellar parameters according to their method they
obtain a scatter diagram with no clear dependence of wind momentum on
luminosity. So in this way we find a conflict between the predictions
of post-AGB evolution theory and the theory of radiatively driven
stellar winds! Are the post-AGB evolutionary tracks not complete? Or
is the behavior of the photon-momentum transfer different in the
atmospheres of O-type CSPNs and massive O-stars?

We now want to rediscuss this situation using our improved model atmospheres.

As a preparatory step we have used our models to calculate the terminal
velocities and mass loss rates for a grid of stars following the current
theoretical post-AGB evolutionary tracks with surface temperatures from 30000 to
90000~K (see, for instance, Bl\"{o}cker~\cite{Bloecker-1995}); the resulting wind
momenta are also plotted in Figure~\ref{fig:wml1} (labelled ``m--L grid''). The
numerical models do nicely follow, as expected, the theoretical wind
momentum--luminosity relation, showing less spread than the ``observed'' values
derived by Kudritzki et~al.~(\cite{Kudritzki-et-al-1997}). The positions of the
Kudritzki et al.\ values in the diagram, compared to those of our models, again
indicate rather large masses between $0.6$ and $0.95\,\Msun$, with a clear
absence of CSPNs with masses below $0.6\,\Msun$.

As explained before, to find so many very massive CSPNs is rather unexpected from
the standpoint of current evolutionary theory, in view of their very high
predicted evolutionary speeds.

\subsection{The relations of the individual dynamical parameters
            $\vinf$ and $\Mdot$}

To try to better understand the discrepancy found from the investigation of the
wind momenta, we must compare the relations of the individual dynamical
quantities involved ($\vinf$ and $\Mdot$), since these relations are not just a
function of the stellar luminosity, as is the case for the
``wind-momentum--luminosity relation'', they are also sensitively dependent on
the stellar mass.

\begin{figure}
\hfil\includegraphics[angle=-90,width=.985\columnwidth]{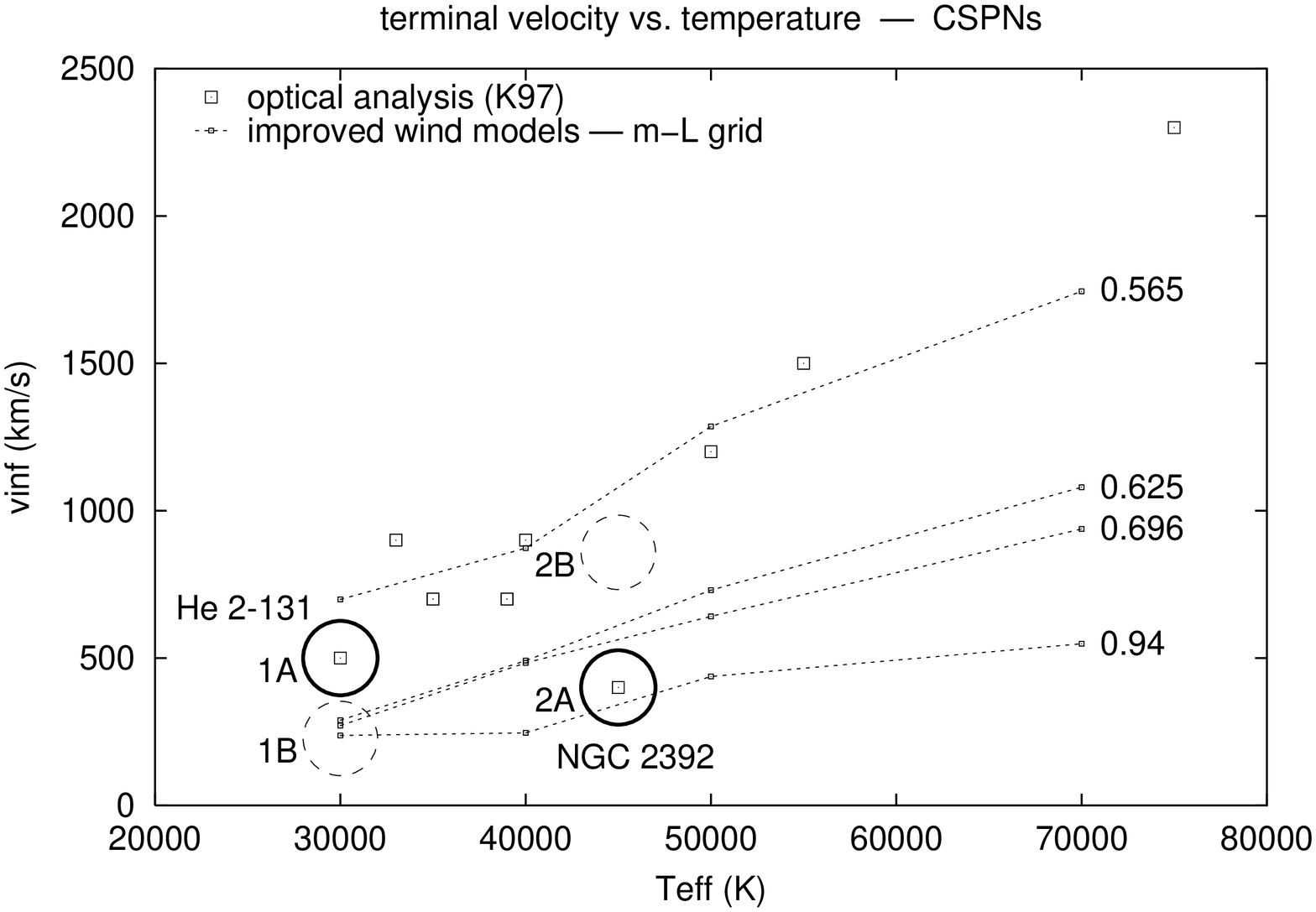}
\includegraphics[angle=-90,width=\columnwidth]{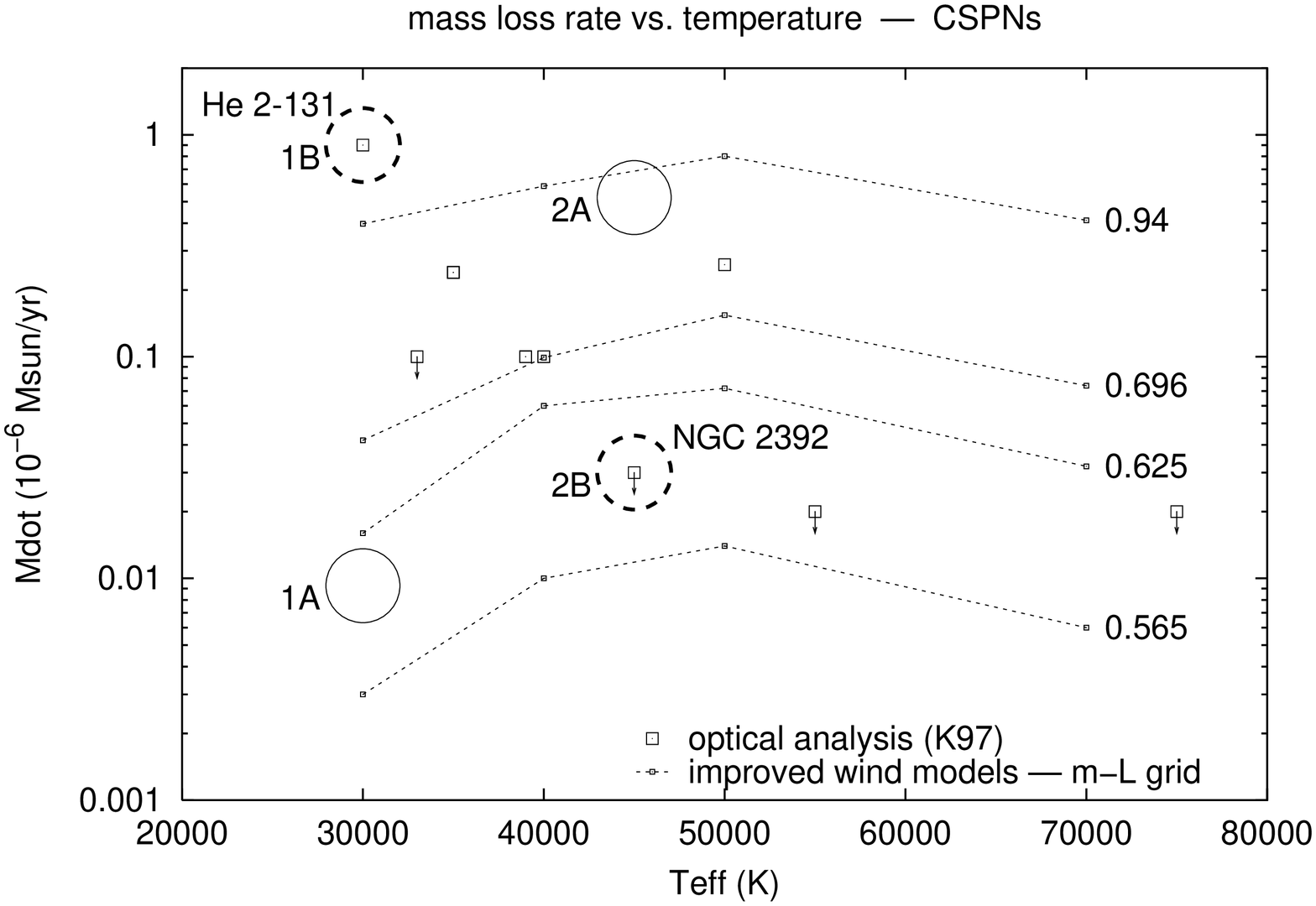}
\caption{
Terminal velocities (upper panel) and mass-loss rates
(lower panel) calculated for a grid of stars following
post-AGB evolutionary tracks (dashed lines, masses in $\Msun$
labelled on the right) compared to observed values derived by
Kudritzki et al.~(\cite{Kudritzki-et-al-1997}) (squares).
Note that the observed terminal velocities
and mass-loss rates indicate different masses
for the same objects (circles) -- for a discussion see text.}
\label{fig:Mdot}
\end{figure}

By doing so we find indeed that something must be seriously wrong.
Figure~\ref{fig:Mdot} (upper panel) shows our predicted terminal velocities and
the observed values. Figure~\ref{fig:Mdot} (lower panel) shows our predicted mass
loss rates and those derived by Kudritzki et al.~(\cite{Kudritzki-et-al-1997})
for their sample. Here a fundamental discrepancy immediately becomes obvious:
whereas the positions of the observations in the diagram showing the terminal
velocities cluster at rather small CSPN masses (between $0.5$ and $0.6\,\Msun$),
their mass loss rates point to a majority of masses above $0.7\,\Msun$.

A detailed look at the positions of individual CSPNs in the plots reveals even
more alarming discrepancies. Take, for example, He~2-131. Its terminal velocity
would indicate a mass of about $0.6\,\Msun$ (circle~1A in Figure~\ref{fig:Mdot}
upper panel). But this mass is completely irreconcilable with its mass loss rate:
it is found not at the position labelled~1A in Figure~\ref{fig:Mdot} (lower
panel), but at~1B, with $\Mdot$ a factor of hundred higher, suggesting a mass of
above $0.94\,\Msun$! The reverse is true for NGC~2392. Its terminal velocity
points to a mass of about $0.9\,\Msun$ (circle~2A in Figure~\ref{fig:Mdot} upper
panel), but its observed mass loss rate is much too small for this mass
(circle~2B in Figure~\ref{fig:Mdot} lower panel), indicating a mass of
approximately $0.6\,\Msun$.

If instead of the terminal velocities we take the mass loss rate determinations
of Kudritzki et al.~(\cite{Kudritzki-et-al-1997}) as basis for the discussion,
then our calculations place these two stars at the positions labelled~1B and~2B
in Figure~\ref{fig:Mdot} (upper panel), with terminal velocities differing by
factors of 2 to~3. But this is clearly ruled out by the observations; $\vinf$ is
a directly measurable quantity!

Therefore, no matter how we look at these plots, we conclude that the
analysis of Kudritzki et al.~(\cite{Kudritzki-et-al-1997}) revealed
mass loss rates which cannot be reconciled with the currently
accepted post-AGB evolutionary tracks which the radiatively driven
wind models shown have been based on.

So far we have used from the UV spectra of the CSPNs only one bit of information:
the observed terminal velocity $\vinf$. Now we will try to clarify the situation
by fitting the full UV spectra with the new atmospheric models. In this way the
wind theory will provide us with stellar parameters derived independently of the
post-AGB evolution theory, and in case we are successful by fitting the spectra
consistently with the dynamical parameters we might have the chance to decide
whether the reason for this discrepancy lies with the evolutionary tracks on the
one hand or the analysis by Kudritzki et al.~(\cite{Kudritzki-et-al-1997}) on the
other.

\section{Consistent UV analysis of the CSPNs He~2-131 and NGC~2392}
\label{sec:cspn-detail}

\begin{figure*}
\includegraphics[angle=90,width=0.49\textwidth]{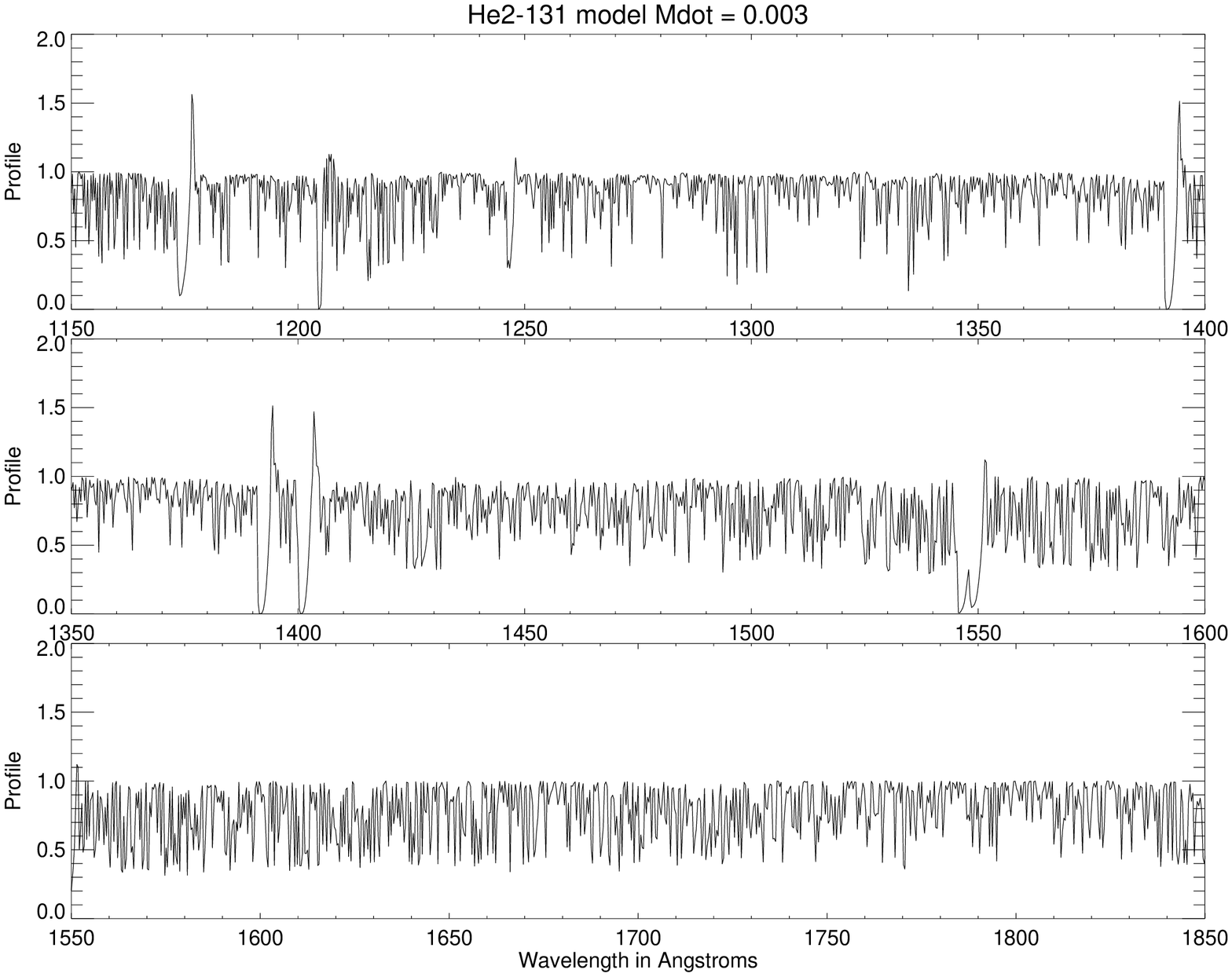}\hfil
\includegraphics[angle=90,width=0.49\textwidth]{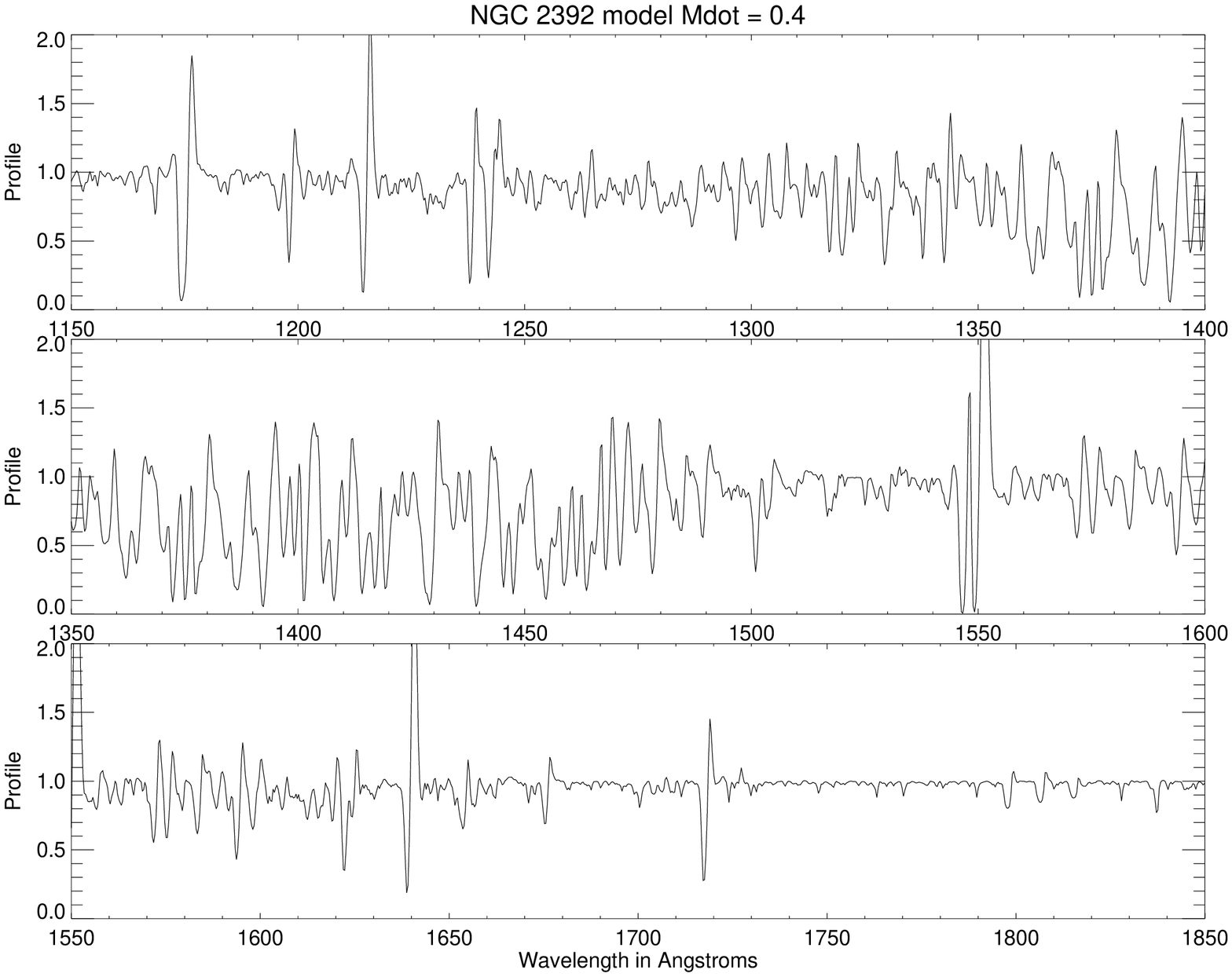}\\[5mm]
\includegraphics[angle=90,width=0.49\textwidth]{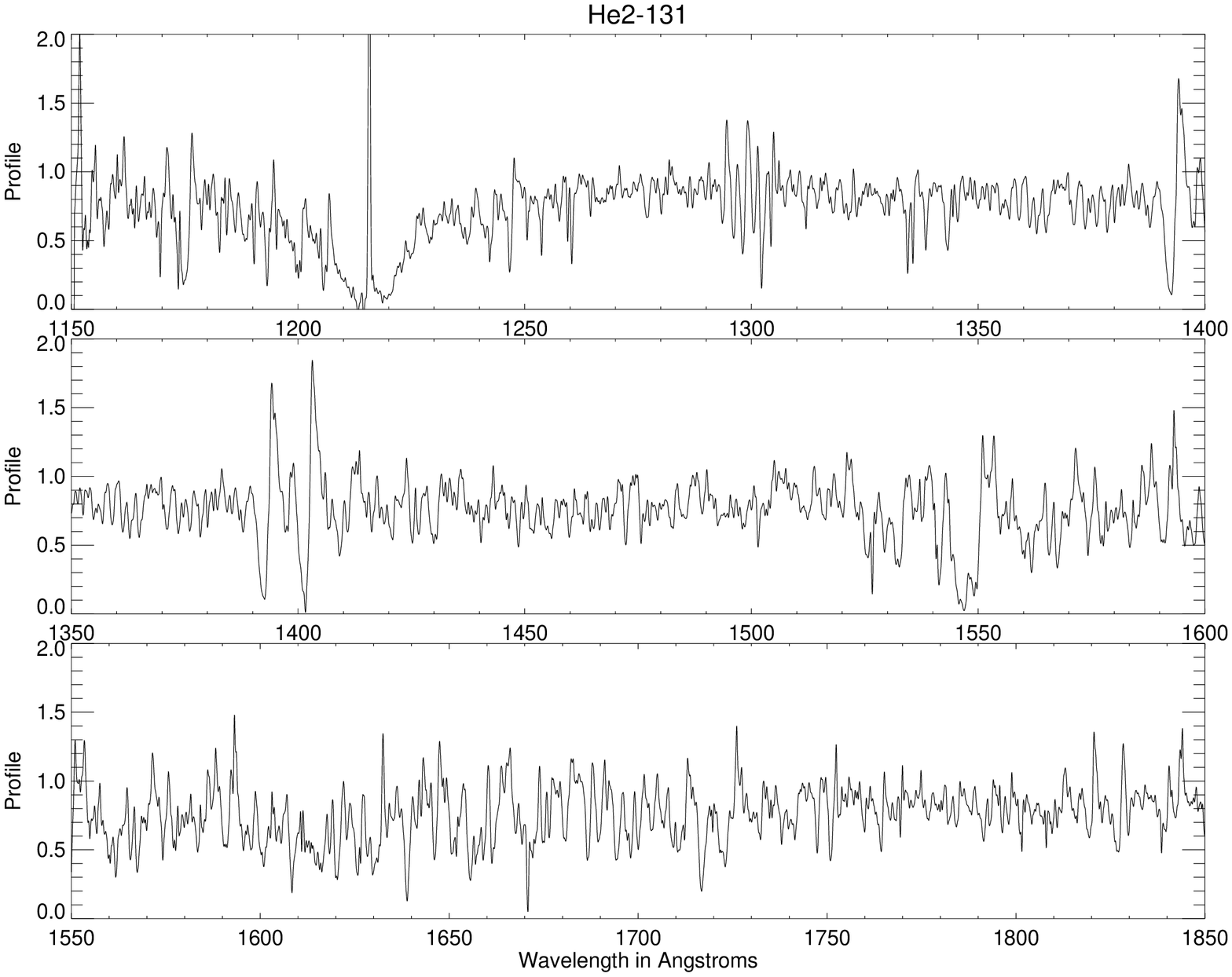}\hfil
\includegraphics[angle=90,width=0.49\textwidth]{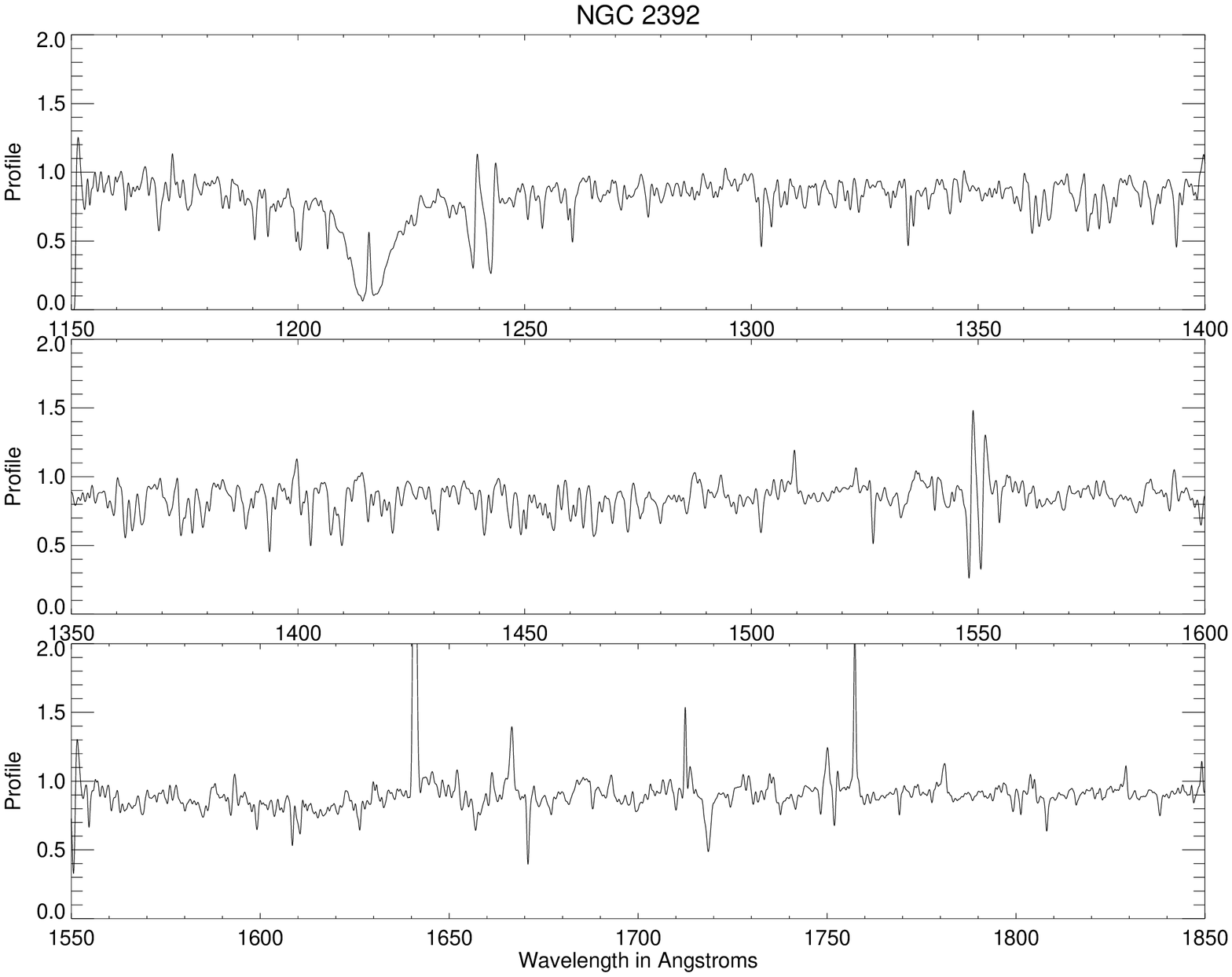}\\[5mm]
\includegraphics[angle=90,width=0.49\textwidth]{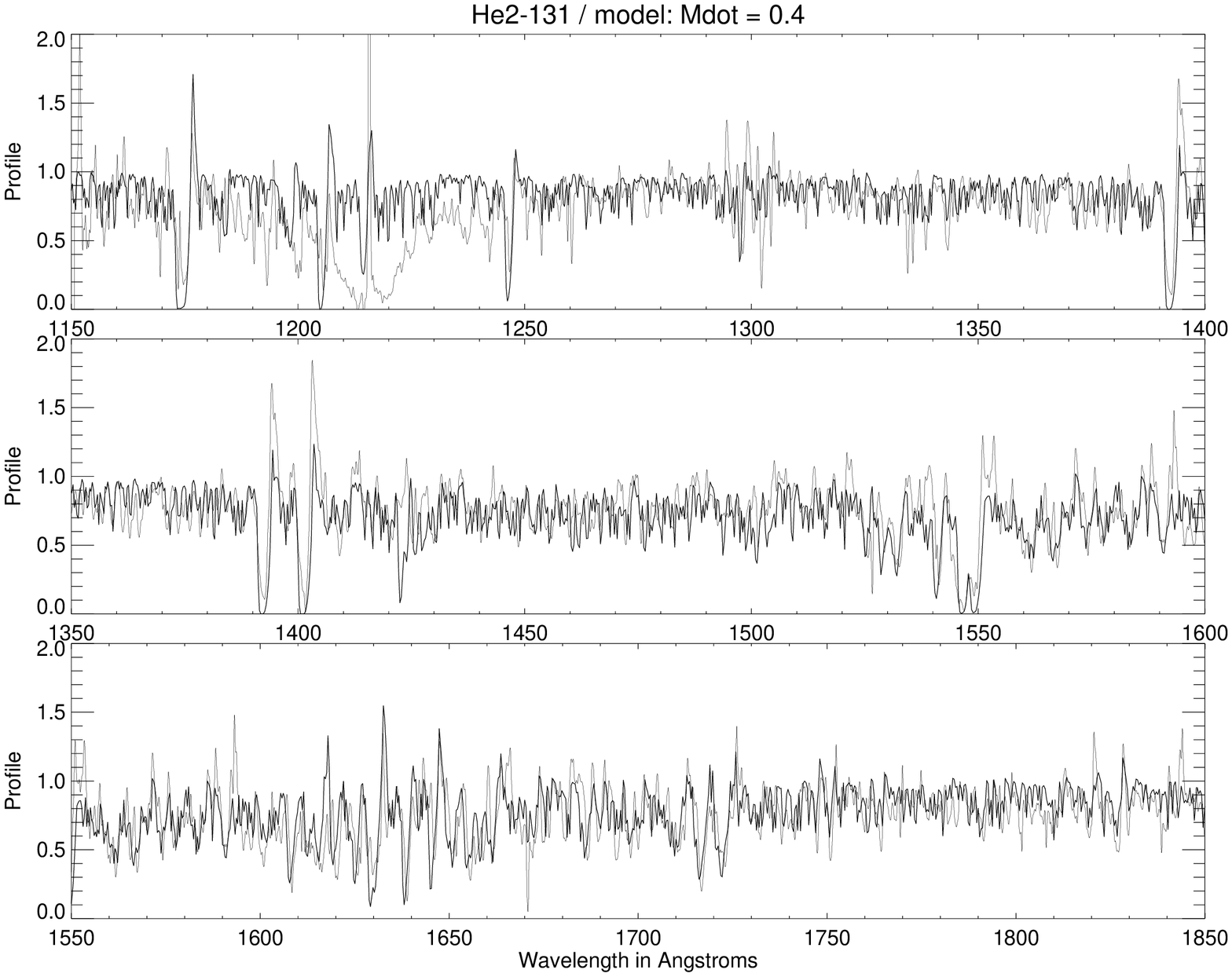}\hfil
\includegraphics[angle=90,width=0.49\textwidth]{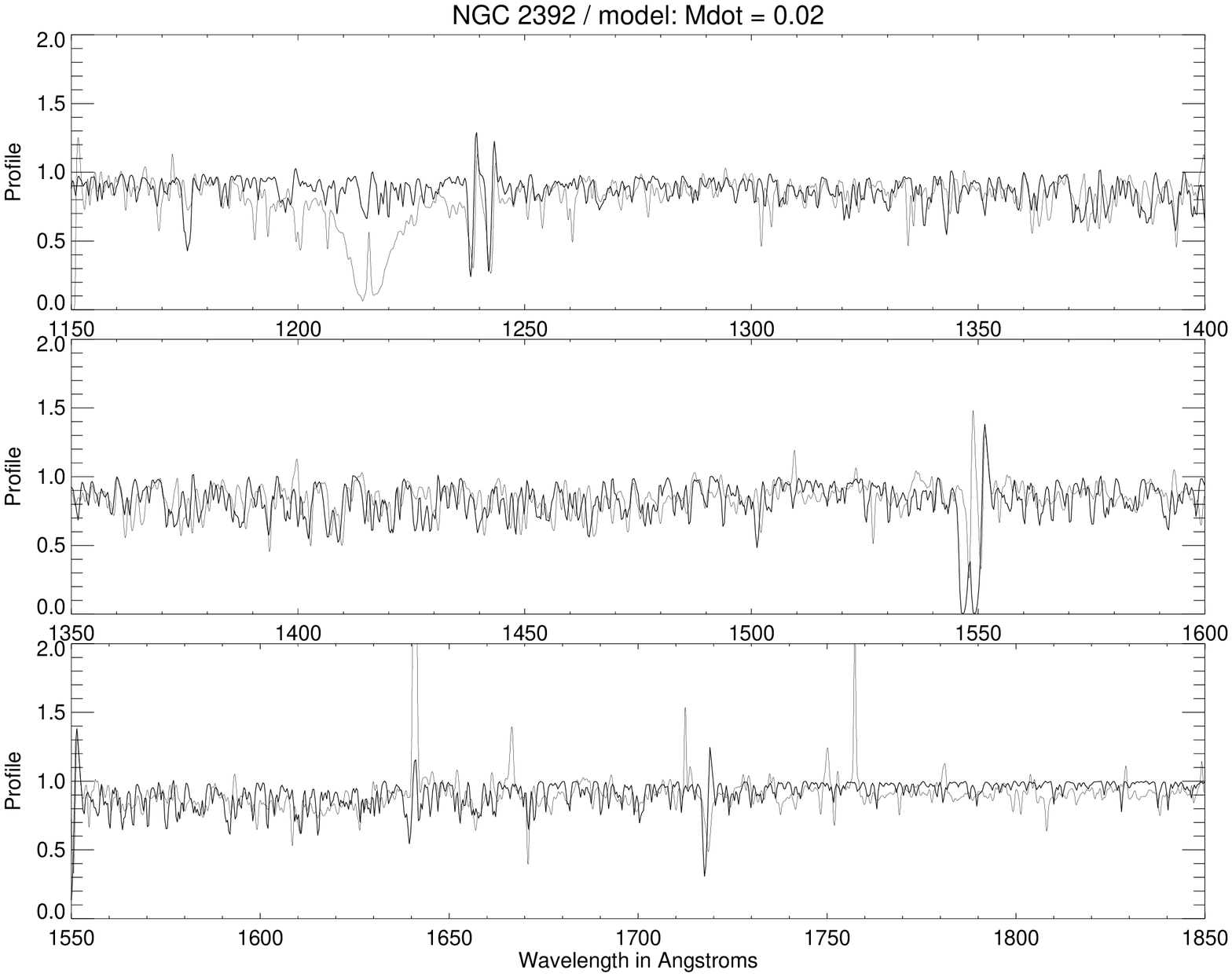}\\
\caption{ {\em (Left)\/}~Top:~Synthetic spectrum of model~1A for \mbox{He~2-131}
(see text). This is incompatible with the observed IUE spectrum (middle). A model
with a significantly enhanced luminosity which gives a higher mass loss rate,
reproduces the distinctive features in the UV spectrum much better (bottom,
overplotted with the observed spectrum to better show the similarity). {\em
(Right)\/}~Top:~Synthetic spectrum of model~2A for NGC~2392. Again, this is
incompatible with the observed IUE spectrum (middle). In this case, however, the
luminosity ($L$) and thus the mass loss rate is much too high, whereas a model
with a lower luminosity reproduces the observed spectrum (bottom).}
\label{fig:spectra}
\end{figure*}

We will start with a detailed description of the two puzzling cases we have been
squarely faced with above.

Figure~\ref{fig:spectra} (top left) shows the synthetic UV spectrum of
the model corresponding to position~1A in Figure~\ref{fig:Mdot}. It
is clearly incompatible with the observed spectrum\footnote{Note
that here and in the following, the observed spectra are contaminated
by interstellar Lyman-$\alpha$ absorption.  We have not attempted
to include this in our models, since the affected region has no
bearing on our conclusions. Neither are other interstellar lines
included in the modelling procedure.} of \mbox{He~2-131} (middle),
since its mass loss rate and due to that its luminosity is obviously
too small, as evidenced by the presence of mostly purely photospheric
lines, hardly affected by the unincisive thin wind. The theory of
radiation-driven winds offers a solution: this CSPN must have a
much larger luminosity, because $L$ is the major factor determining
the mass loss rate.  We have calculated a series of models with
increasing luminosity -- and therefore increasing mass loss rate --
(at the same time adjusting the mass to keep the terminal velocity at
its observed value) to verify if one of these models could reproduce
the numerous strongly wind-contaminated iron lines observed especially
between $1500$ and $1700$\,\AA. Indeed, a more luminous model, able to
sustain the high mass loss rate of model~1B in Figure~\ref{fig:Mdot},
produces a convincing fit -- see Figure~\ref{fig:spectra}, bottom left.
The parameters of this model are given in Table~\ref{tbl:params}.

The situation is reversed with NGC~2392. The synthetic spectrum of model~2A in
Figure~\ref{fig:Mdot} is incompatible with the observed UV spectrum
(Figure~\ref{fig:spectra}, right top and middle, respectively), since it produces
many strongly wind-contaminated lines, which are not observed; the star produces
almost exclusively photospheric lines! Again the problem can be attributed to the
luminosity, which is too high in this case. Decreasing the luminosity and thus
the mass loss rate yields a model which is quite well in agreement with the
observed spectrum (Figure~\ref{fig:spectra}, bottom right). The stellar
parameters of this model are also given in Table~\ref{tbl:params}.

In summary, the new model atmospheres produce a good fit to all
the observable features in the UV spectrum.  We remark at this
point that our error in the stellar mass is very small \mbox{($\le
0.1\,\Msun$)} due to the sensitive dependence on the predicted
$\vinf$ and the small error received from determining this value
from the observed spectrum ($\le 10\%$). Furthermore, we note that
our predicted values of $\vinf$ are in agreement within 10\% with
the observed values in the case of massive O~stars (cf.~Hoffmann and
Pauldrach~\cite{Hoffmann-Pauldrach-2001}). Thus, a possible internal
error leaves no margin for a larger uncertainty in the deduced masses.

What can we conclude from the derived stellar parameters? Let us
consider first the weak-winded CSPN, NGC~2392. We determine a $\Teff$
of 40000~K from the ionization equilibrium of Fe ions in the stellar
UV spectrum, similar to the value obtained from the ionization
equilibrium of He\,{\sc i} and He\,{\sc ii} (absorption lines in
the optical stellar spectrum).\footnote{We do not want to hide the
fact that this central star has an anomalously high He\,{\sc ii}
Zanstra temperature of about 70000~K and an even higher energy-balance
temperature (M\'endez et al.~\cite{Mendez-et-al-1992}), but we have
carefully verified that both the visual and especially the UV stellar
features are decidedly incompatible with such high temperatures.
This discrepancy is at present unresolved. Apparently an additional
source of He-ionizing photons is needed in this case. For the moment
we ignore this, and perform the analysis using the information about
$\Teff$ derived from the stellar spectrum.} The very low terminal
wind velocity of $400\,{\rm km\,s}^{-1}$, together with the low
luminosity (needed to adjust the predicted mass loss rate so that
the predicted and observed spectra are in good agreement) lead us to
adopt a small radius. Using this radius ($1.5\,\Rsun$) and $\vinf$
we get a stellar mass of only $0.41\,\Msun$, a value much smaller than
obtained if we assume the classical post-AGB mass--luminosity relation
-- a high mass of $0.9\,\Msun$ was the result found by Kudritzki et
al.~(\cite{Kudritzki-et-al-1997}).


Note that the radius $1.5\,\Rsun$ and mass $0.41\,\Msun$ of
this CSPN correspond to $\logg = 3.7$, in good agreement with the
$\logg$ derived earlier from the NLTE plane-parallel analysis of the
optical stellar spectrum.

In the case of the central star of He~2-131 the terminal velocity of $500\,{\rm
km\,s}^{-1}$ (and $\Teff=33000\,{\rm K}$) would appear to suggest, according to
the classical post-AGB mass--luminosity relation (cf.~Figure~\ref{fig:Mdot} upper
panel), a stellar mass of about $0.6\,\Msun$. However, the wind features observed
in the UV spectrum forced us to increase the stellar $R$ and thus $L$, which in
turn increased $\Mdot$ until a good fit was obtained. From the corresponding
large radius -- $5.5\,\Rsun$ -- and $\vinf$ we derive a stellar mass of
$1.39\,\Msun$, a value very close to the Chandrasekhar mass limit for white
dwarfs! Thus, in this case the resulting mass is even more extreme than the value
of $0.9\,\Msun$ obtained by Kudritzki et al.~(\cite{Kudritzki-et-al-1997}).

\section{Consistent UV analysis of 7 additional CSPNs}
\label{sec:cspn-rest}

\begin{figure*}
\includegraphics[angle=90,width=0.49\textwidth]{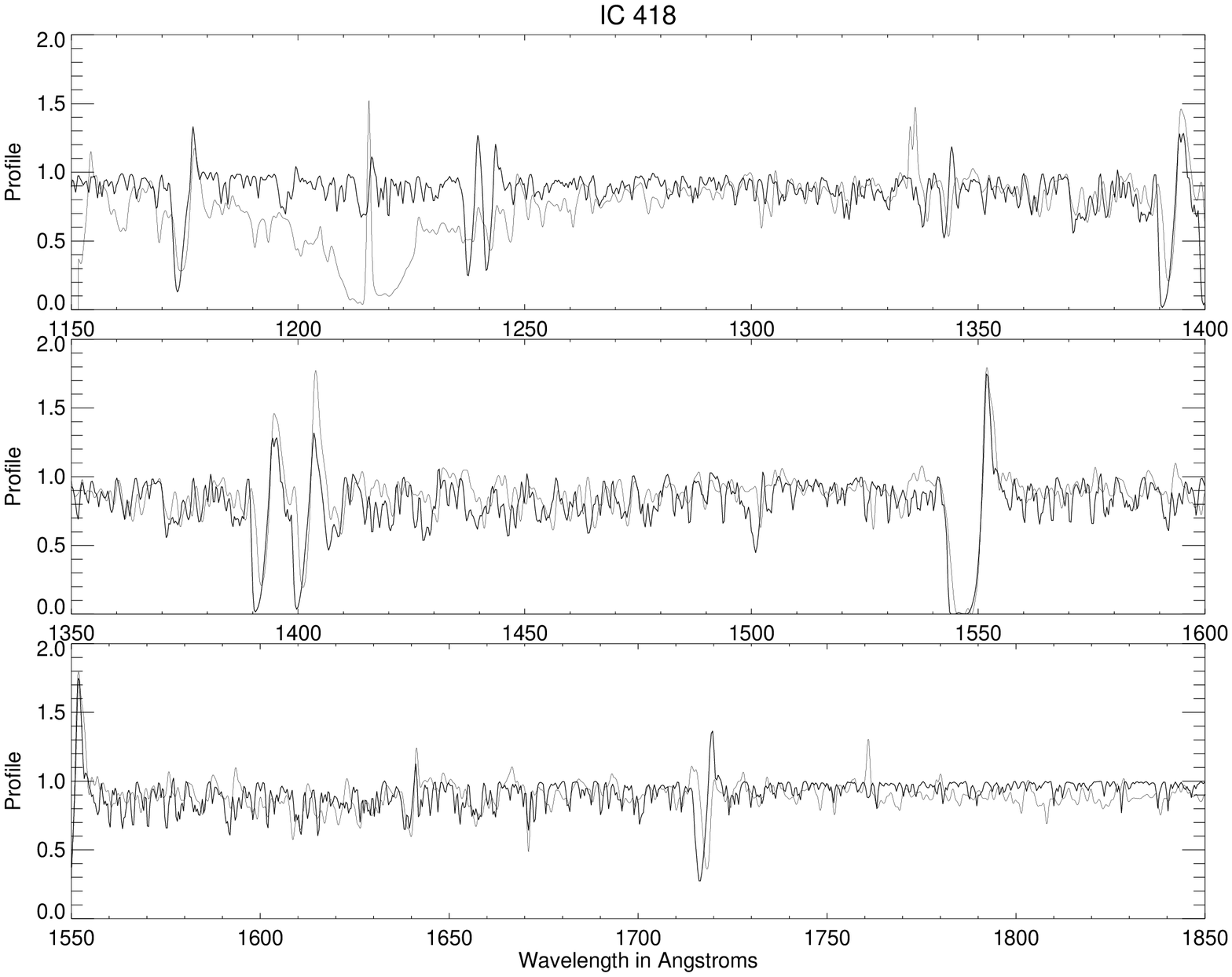}\hfil
\includegraphics[angle=90,width=0.49\textwidth]{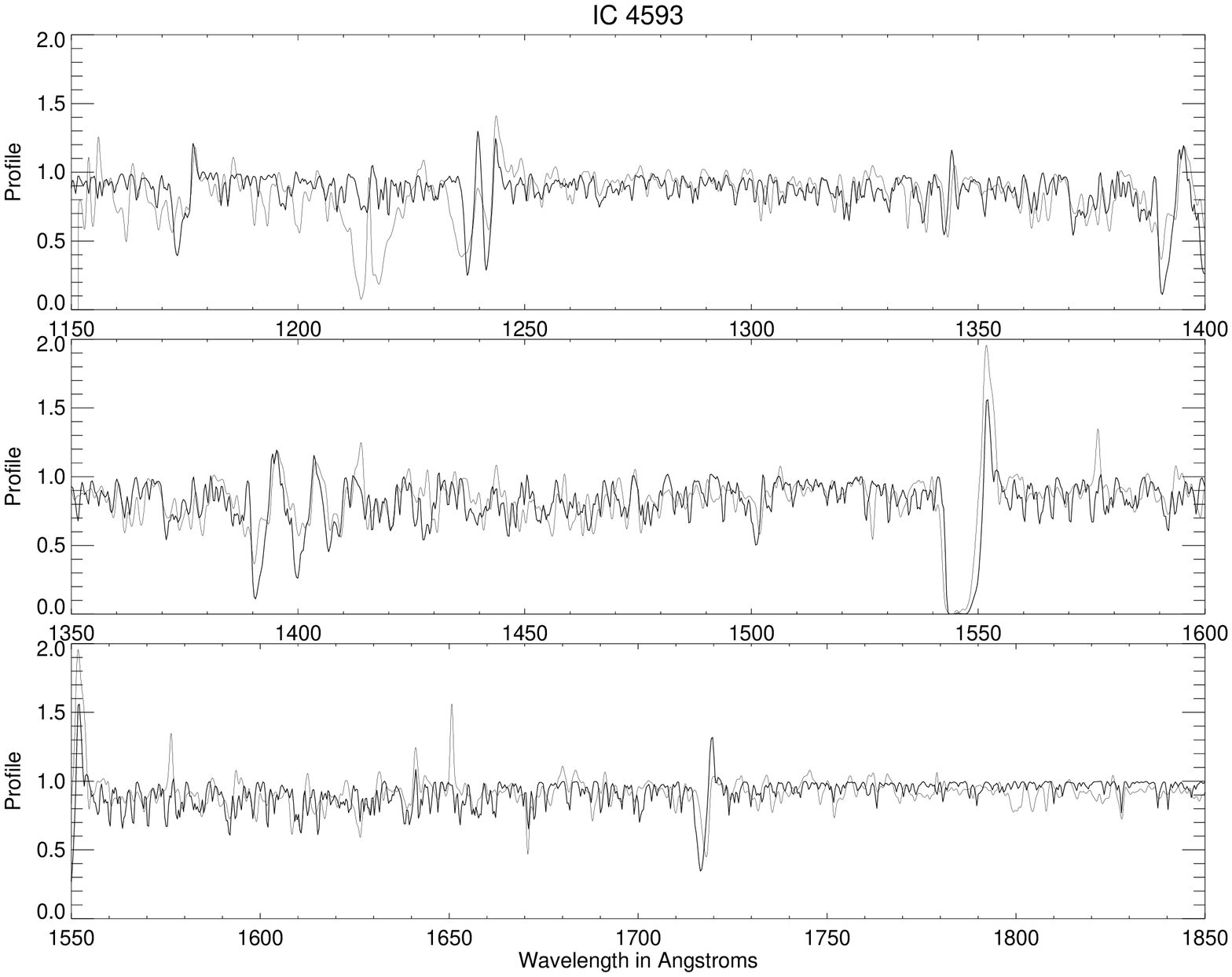}\\[5mm]
\includegraphics[angle=90,width=0.49\textwidth]{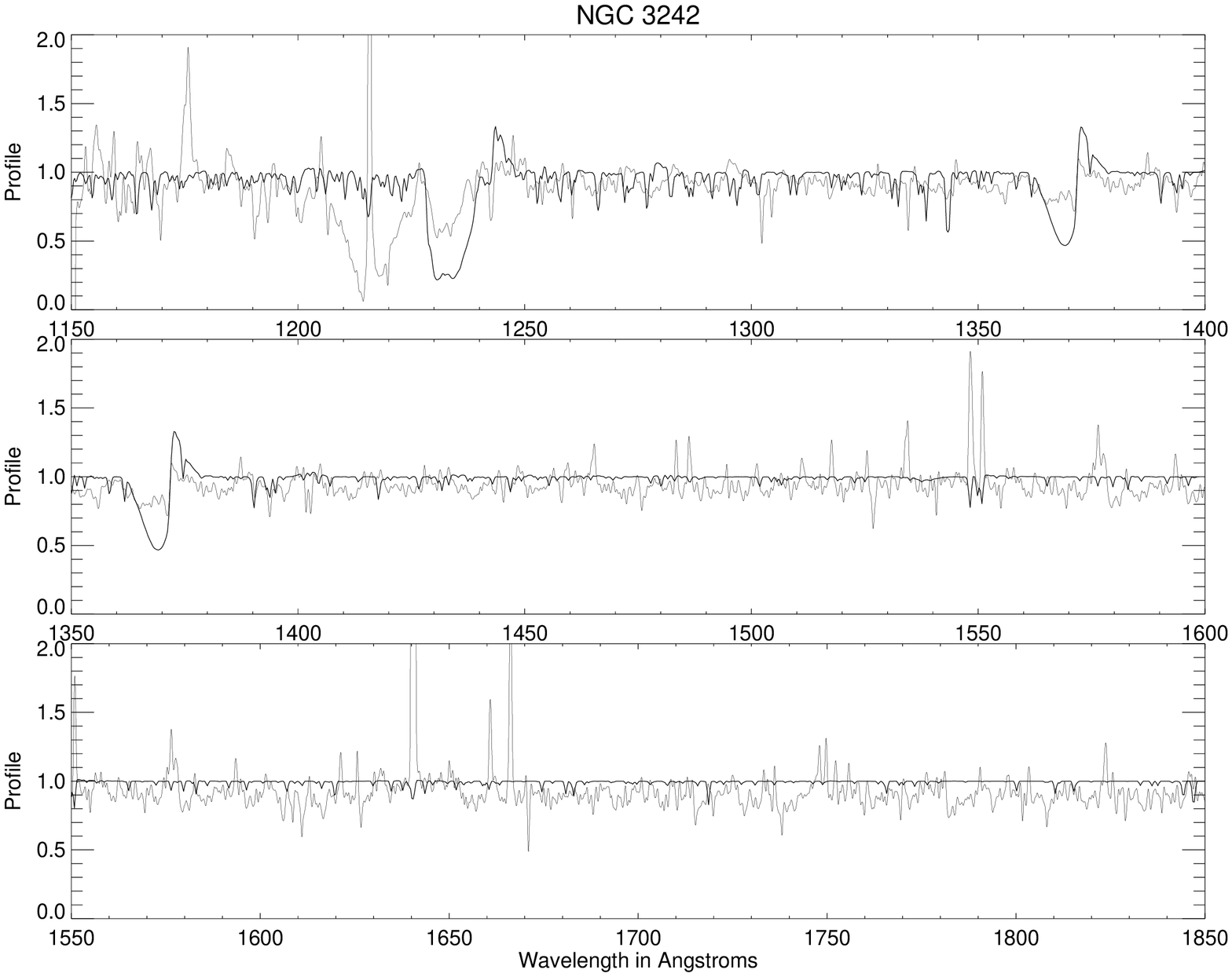}\hfil
\includegraphics[angle=90,width=0.49\textwidth]{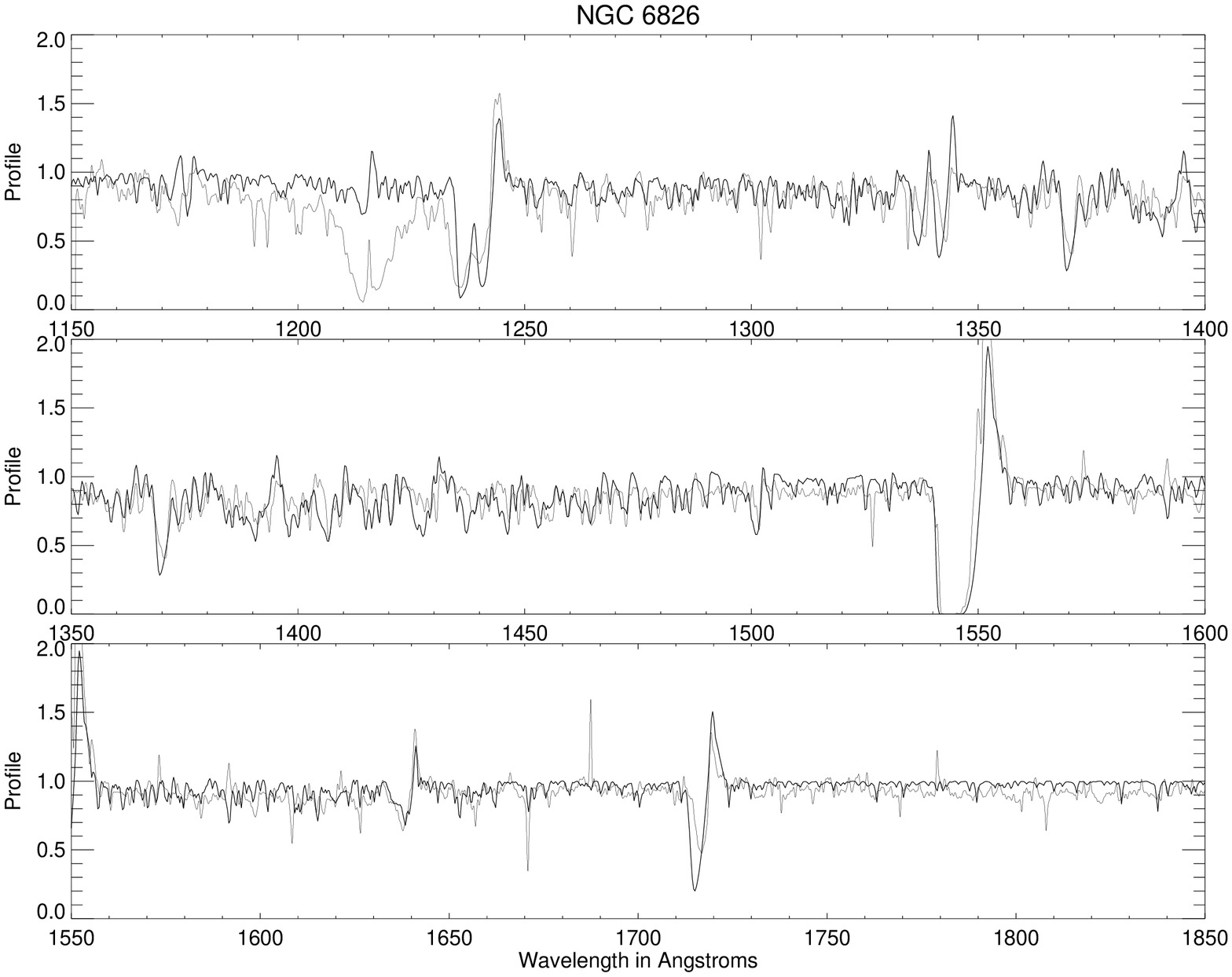}\\[5mm]
\includegraphics[angle=90,width=0.49\textwidth]{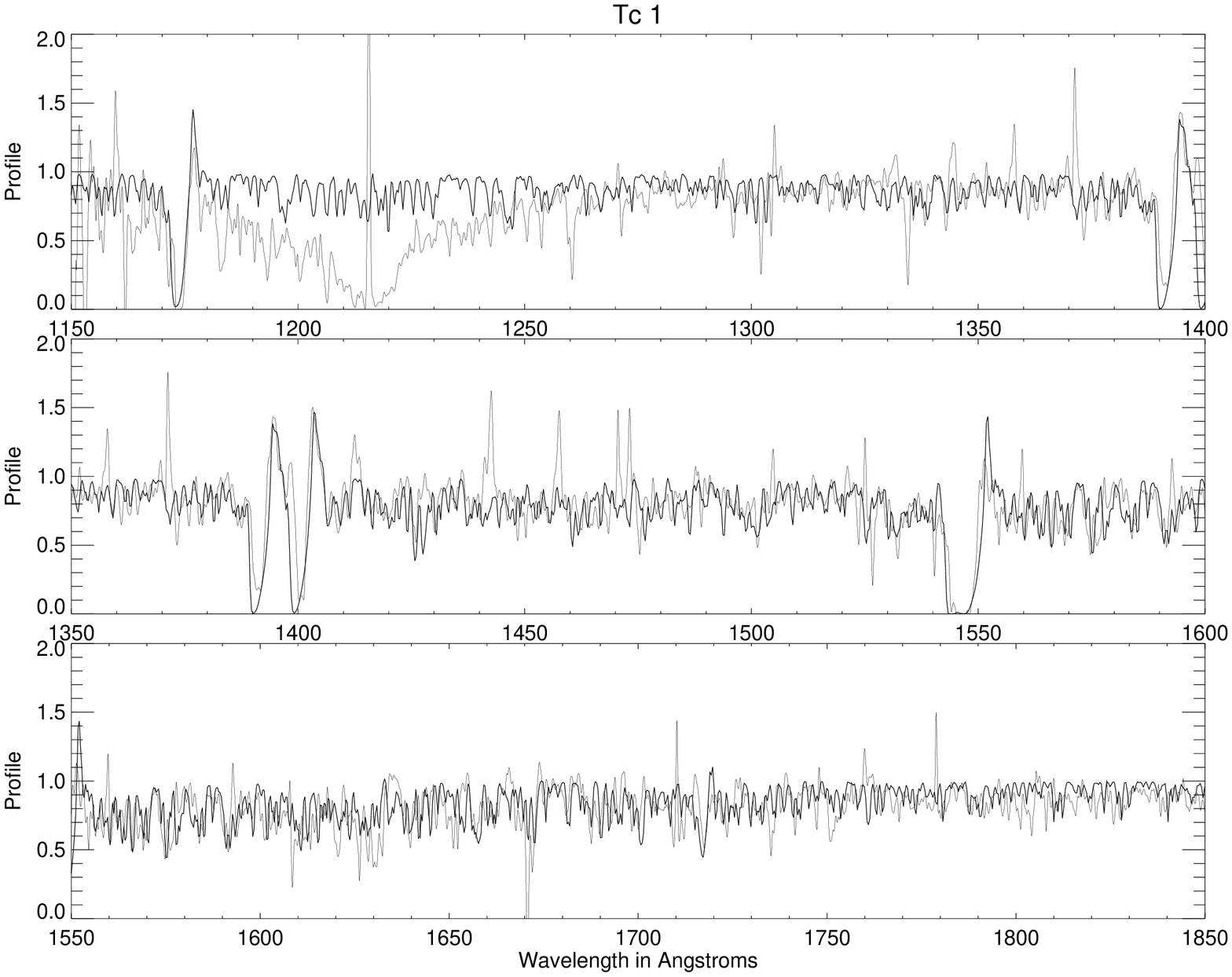}\hfil
\includegraphics[angle=90,width=0.49\textwidth]{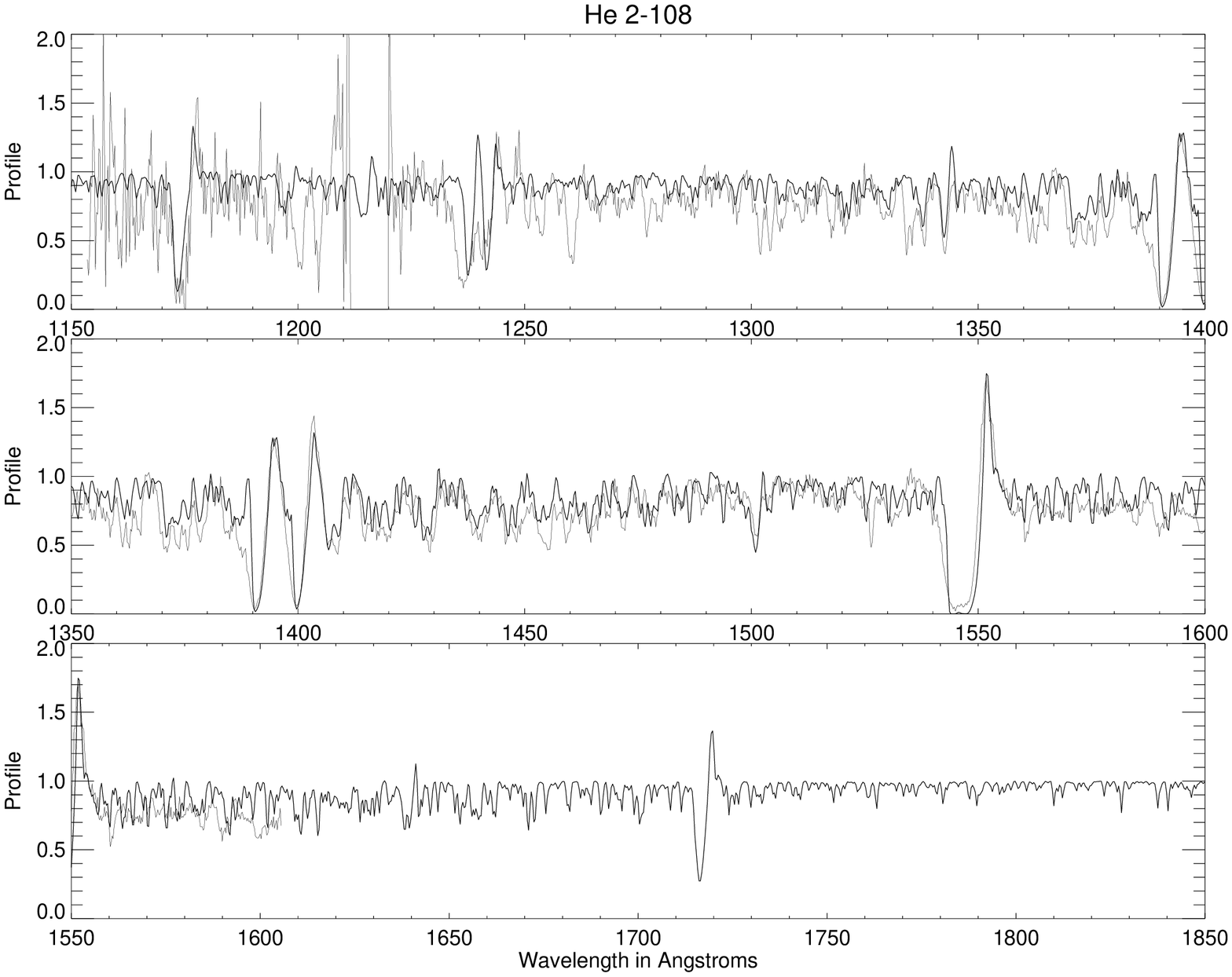}\\
\caption{ Synthetic UV~spectra from consistent atmospherical models for the other
6 CSPNs of our sample, compared to the observed UV~spectra: IC~418 and IC~4593
{\em (top)}, NGC~3242 and NGC~6826 {\em (middle)}, and Tc~1 and He~2-108 {\em
(bottom)}. } \label{fig:otherspectra}
\end{figure*}

In a similar fashion as for the two objects described in detail
in the previous section, we have computed hydrodynamical models
that reproduce the observed UV~spectra of the 7 other CSPNs of our
sample. The spectra are plotted in Figure~\ref{fig:otherspectra},
the resulting parameters are listed in Table~\ref{tbl:params}.

We would like to remark on two points in this context.  The first
is that the observed spectrum of IC~4637 (not shown) is very noisy.
Our parameters given for this particular star are therefore not of
the same quality as those of the other objects, and should thus be
seen more as a hydrodynamic consistency check to the values derived
by Kudritzki et al.~(\cite{Kudritzki-et-al-1997}), than as resulting
from a detailed spectral fit.

The second is that the UV~spectra of IC~418 and He~2-108 are very
similar, and we therefore derive similar parameters for these two
stars, as the same model obviously fits both spectra equally well.

Concerning the elemental composition, we have adopted the Helium
abundances from Kudritzki et al.~(\cite{Kudritzki-et-al-1997}).
For the other elements, we have assumed a solar abundance pattern,
justified by the good fit to the observed UV~spectra. A minor
discrepancy is seen in NGC~2392, which the optical spectrum would
indicate to be N-rich and C-deficient, a result also reflected in the
UV~spectrum (for example, both the observed C\,{\sc iii} and C\,{\sc
iv} lines are weaker than those of the model). However, the influence
of this on the hydrodynamics is small, since C and N are not major
contributors to the line force (Pauldrach~\cite{Pauldrach-1987})
and the sum of C, N, and O would remain constant if these abundances
were the result of the CNO-process.

\section{Interpretation of CSPN winds}
\label{sec:interpretation}

\begin{table*}
\caption{Parameters of nine CSPNs derived from our analysis of
the UV spectra using our model atmospheres,
compared to the values found by Kudritzki et
al.~\cite{Kudritzki-et-al-1997}.}
\label{tbl:params}
\begin{center} \input{table1.tex} \end{center}
\end{table*}

Table~\ref{tbl:params} shows the result of applying the
method and UV analysis to the nine CSPNs of our sample.


First of all, these results are, in a broad sense, encouraging.
We have not found any object with decidedly impossible
masses and luminosities (for example, we could have derived masses
and luminosities typical of massive Pop.\,I stars: certainly that
would have been quite embarrassing, but it did not happen).
However, a closer look shows that we are in a very unexpected
situation.

\begin{figure}
\includegraphics[angle=-90,width=\columnwidth]{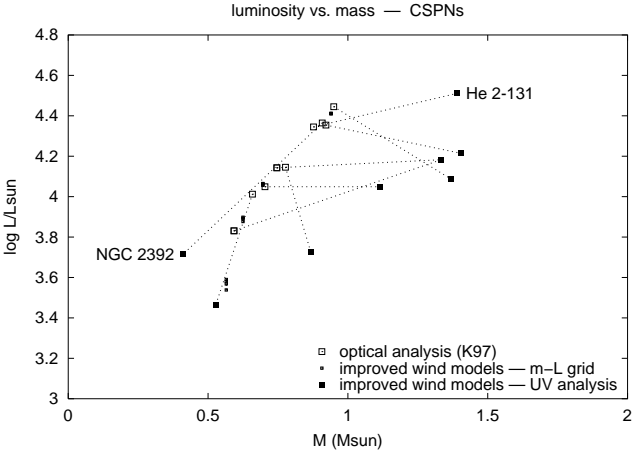}
\caption{
Luminosity vs.\ mass for the evolutionary tracks (open squares)
compared to the observed quantities determined with our method
(filled squares).
Although the luminosities deduced from the UV spectra lie in
the expected range, a much larger spread in the masses
(from $0.4$ to $1.4\,\Msun$) is obtained.
The relation between CSPN mass and luminosity deviates severely from
that taken from the theory of post-AGB evolution.
Of course the latter is followed by the open squares and dots,
because in that case it had been assumed from the start.}
\label{fig:ml}
\end{figure}

Figure~\ref{fig:ml} shows the relation between stellar mass
and luminosity obtained from our model atmosphere analyses,
in comparison with the mass--luminosity relation of the
evolutionary tracks, represented by the values from Kudritzki et
al.~(\cite{Kudritzki-et-al-1997}). From the viewpoint of current
stellar evolutionary calculations this plot is somewhat unsettling:
there is a very large spread in masses, between 0.4 and
1.4\,$\Msun$, and the derived masses and luminosities do not agree
with the classical post-AGB mass--luminosity relation. Most CSPNs are
underluminous for their mass (or too massive for their luminosities).

\begin{figure}
\includegraphics[angle=-90,width=\columnwidth]{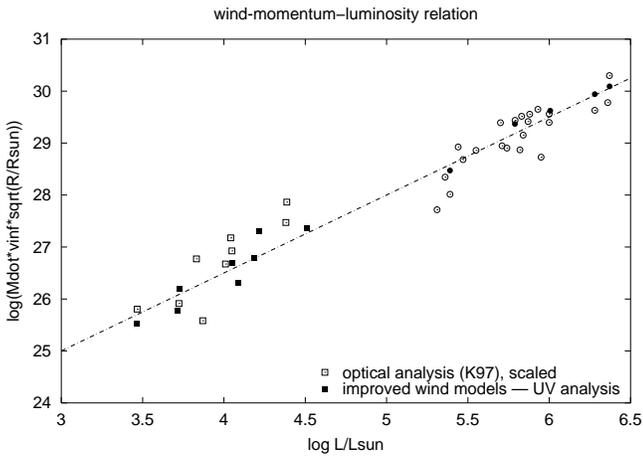}
\caption{ The wind-mo\-mentum--luminosity relation for CSPNs (lower left) based
on our values determined from the UV spectra (filled squares). The open squares
are the values from Kudritzki et al.~(\cite{Kudritzki-et-al-1997}) scaled to our
determined radii, thus eliminating the need of an a-priori assumption for the
radii, as was done by Kudritzki et al. (see text). Compared to
Figure~\ref{fig:wml1} the result is striking: now there is good agreement between
the optical and UV mass loss rate determinations.} \label{fig:wml2}
\end{figure}

In Figure~\ref{fig:wml2} we show again the wind-momentum--lumi\-nos\-ity relation
for both massive hot stars and CSPNs, but this time based on the parameters
derived in our analysis. Our new parameters give wind momenta of the right order
of magnitude and within the expected luminosity range (there may be still too
many CSPNs at $\log L/\Lsun > 4$, but not so many as in Kudritzki et
al.~\cite{Kudritzki-et-al-1997}). The CSPNs are found along the extrapolation of
the wind-momentum--luminosity relation defined by the massive hot stars, and the
CSPNs show a smaller dispersion, i.e., a tighter correlation of wind-momentum
with luminosity, than was the case in Kudritzki et
al.~(\cite{Kudritzki-et-al-1997}). None of these facts is surprising, because our
derived parameters are now based on the wind theory; of course the theory,
consistently applied, will not produce any departure from its own predictions!
However, the really significant fact is that we could produce a very convincing
fit simultaneously to a multitude of diagnostic features in the CSPN UV spectra.
There was no guarantee a priori that such a good overall fit was possible, and
this is the main reason why we think that it is not easy to simply argue ``the
wind models must be wrong''. Instead, it is very likely that the theory and the
models as an approach to it are correct in case the experiment in the form of a
comparison of observed and synthetic UV-spectra was successful.

More importantly, the results obtained for NGC~2392 and NGC~3242 rule out the
possibility that our method simply systematically overestimates the stellar
masses: for these two stars we derive masses that lie below those deduced by
Kudritzki et al.~(\cite{Kudritzki-et-al-1997}). In the extreme case of NGC~2392,
any systematic overestimate of the mass would require this star to have a mass
even below $0.4\,\Msun$. If there is any physical effect at work whose neglect
results in a systematic error in the analysis, it would need to be such that it
can lead to an over- as well as underestimate of the masses, despite reproducing
nearly perfectly the observed UV~spectra. Our current knowledge of stellar winds
does not provide us with any mechanism able to do this.

If we drop the {\em assumption\/} made by Kudritzki et~al.\ that the
stars obey the theoretical post-AGB mass--luminosity relation, and
instead scale their mass loss rates to our radii\footnote{Additionally
allowing for their different effective temperatures by requiring
that the observed visual flux ($\sim R^2 \Teff$) stay constant.}
-- keeping $Q$, the real observational quantity, fixed -- then
their wind momenta match ours to within about a factor of two.
Furthermore, their sample with the radii thus scaled now also shows
a much tighter correlation of the wind momentum to luminosity than
before (see Figure~\ref{fig:wml2}). This indicates that {\bf two
independent procedures\/} to obtain the mass loss rates (one based
on optical, the other based on UV wind-sensitive line profiles) {\bf
have given consistent results}. In other words, the problem cannot
be attributed exclusively to the observational data used by Kudritzki
et al.~(\cite{Kudritzki-et-al-1997}) to estimate the mass loss rates.

If we believe both, the current evolutionary theory and the luminosities and
masses we have determined from the atmospheric models, then most of our CSPNs
have not followed a classical post-AGB evolution. We find many stars near the
Chandrasekhar limit for white dwarfs. They do not obey the core-mass--luminosity
relation (being underluminous for their mass) and this indicates that their
internal structure must be different. The special case of NGC~2392 is also
remarkable: with such a small mass it cannot be a post-AGB star, and we would be
forced to consider alternative evolutionary histories, involving probably a
binary system merged immediately after the first visit to the red giant branch. A
few similar cases of low-mass CSPNs have been noted in the past: EGB~5 and
PHL~932, see M\'endez et al.~(\cite{Mendez-et-al-1988a},
\cite{Mendez-et-al-1988b}). What makes NGC~2392 a more troublesome case is the
additional fact that kinematically it is a rather young PN, while numerical
simulations of binary merging lead to expect no visible nebulae around them, or
at most very old ones, like EGB~5 and PHL~932.

If we reject such drastic departures from the classical post-AGB evolutionary
picture, still assuming the evolutionary calculations to be correct, then we
would need to conclude that our models, while adequate for massive O~supergiants,
are a failure for stars of similar surface temperature and gravity in another
evolutionary status, and produce good fits to the CSPN UV~spectra only by a
surprising and misleading coincidence.  Considering the successes of
radiatively-driven wind theory, however, we regard this conclusion as highly
improbable.  We must therefore contemplate the possibility that our current
knowledge of stellar evolution might be incomplete.

\section{Spectroscopic distances and white dwarf masses}
\label{sec:other-observations}

\begin{table*}
\caption{Computed spectroscopic distances of our sample stars and the
quantities used to derive them. (See text.)}
\label{tbl:distances}
\begin{center} \input{table2.tex} \end{center}
\end{table*}

Facing this surprising situation, we ask if there is any way of further verifying
the CSPN masses and luminosities we determined. One possible way is to calculate
the spectroscopic distances and verify if they agree with the rest of the
available evidence. Another way, since we expect CSPNs to become white dwarfs, is
to look into what is currently known about the white dwarf mass distribution and
into the recent results on asteroseismology of pre-white dwarfs.

Having all the basic stellar parameters it is easy to calculate the
spectroscopic distances, following, for example, the method described
in M\'endez et al.~(\cite{Mendez-et-al-1992}), which uses the stellar
mass, $\logg$, monochromatic model atmosphere flux at visual wavelength,
and dereddened apparent visual magnitude.

Table~\ref{tbl:distances} shows these quantities and the
resulting distances. They are not too different from the earlier
spectroscopic distances by M\'endez et al.~(\cite{Mendez-et-al-1988b},
\cite{Mendez-et-al-1992}), except for the effect of the different
stellar masses we are using now.  NGC~2392 gets a smaller distance
and He~2-131 a larger one because the new mass is smaller and larger,
respectively, than before.

Now we want to discuss systematically how these spectroscopic
distances compare with individual PN distances derived from other
methods. We disregard all the statistical distances published by
many authors because they are too uncertain for a case-by-case
discussion. We consider in turn trigonometric parallaxes,
distances derived from visual companions to the central stars,
cluster distances, extinction distances, and expansion distances.

\subsection{Trigonometric parallaxes}

There is no overlap with the sample we are analyzing here. However,
it is possible to compare trigonometric versus other spectroscopic
parallaxes. Some ground-based trigonometric parallaxes are listed
by Harris et al.~(\cite{Harris-et-al-1997}). We consider only those
objects with reliable parallaxes larger than 2~milli-arc-seconds,
i.e., with distances below 500~pc: NGC~6853, NGC~7293, S~216. Their
distances compare very well with the spectroscopic distances of
Napiwotzki~(\cite{Napiwotzki-1999}).

The Hipparcos parallaxes again do not overlap with our sample,
but again it is possible to compare with other spectroscopic
parallaxes. Here there is some disagreement, especially in the
case of PHL~932 and (marginally) A~36. The parallax for NGC~1360 is
too uncertain, but Pottasch and Acker~(\cite{Pottasch-Acker-1998})
show convincingly that the Hipparcos distances of PHL~932 and A~36
require higher surface gravities than indicated by the spectroscopic
analysis {\em if we assume a central star mass of} 0.6~$\Msun$. One
way to reduce the discrepancy is to reduce the mass of the central
star; so in fact we could argue that Hipparcos has confirmed the
conclusion that the central star of PHL~932 must have a very low
mass, below 0.3~$\Msun$, and cannot be a post-AGB star (M\'endez et
al.~\cite{Mendez-et-al-1988a}). Napiwotzki~(\cite{Napiwotzki-1999})
has repeated the spectroscopic analysis of PHL~932, using different
models and spectrograms, and obtains atmospheric parameters marginally
consistent with those of M\'endez et al.~(\cite{Mendez-et-al-1988a}).
His surface gravity is somewhat higher, but not as high as required
by the Hipparcos parallax, unless we reduce the central star mass to
the rather implausible value of 0.1~$\Msun$. Therefore in the case of
PHL~932 some degree of contradiction remains. Would somebody please
remeasure this parallax? Nowadays it can probably be done from the
ground with adequate CCD techniques.

\subsection{Distances derived from visual companions to the CSPNs}

Again no overlap; but Ciardullo et al.~(\cite{Ciardullo-et-al-1999})
assign a distance of 2.4~kpc to NGC~1535, in good agreement
with the spectroscopic distance of 2.0~kpc in M\'endez et
al.~(\cite{Mendez-et-al-1992}).

\subsection{Cluster distances}

Again no overlap. The only object we can mention here is the central
star of the PN in the globular cluster M~15, where the spectroscopic
distance is in excellent agreement with the cluster distance, see
McCarthy et al.~(\cite{McCarthy-et-al-1997}).

\subsection{Extinction distances}

Here finally we find some objects in common with our sample.
Martin~(\cite{Martin-1994}) concludes that the extinction distance
of He~2-131 (about 700~pc), although substantially smaller,
does not necessarily invalidate our distance, in view of the high
Galactic latitude of this PN. The same point was made earlier by
Maciel~(\cite{Maciel-1985}): given its high latitude, this object at a
distance of 700~pc would be some 180~pc below the Galactic plane, which
is not very different from the halfthickness of the Galactic absorbing
layer. As a consequence, the extinction distance should be taken as a
lower limit to the true distance. The same can be argued about Tc~1,
with an extinction distance of 600~pc, although Martin considers this
case more of a contradiction with the spectroscopic distance. For
another of our objects (He~2-108) again there is no conflict.

Martin's best case of a contradiction is IC~2448, with an
extinction distance of 840~pc and a spectroscopic distance of about
3~kpc. Unfortunately IC~2448 is not in the sample we are studying
here; we do not know if a revised spectral analysis would reduce
its spectroscopic distance somewhat. In any case, we think that one
isolated discrepant case does not have too much weight, given the
existence of many other cases showing agreement, because the isolated
discrepancy can always be attributed to accidental fine structure in
the interstellar dust distribution.

\subsection{Expansion distances}

Again no overlap with our sample, but since we mentioned
IC~2448 in the previous subsection, we should add that there
is a recent expansion distance estimate for this PN by Palen et
al.~(\cite{Palen-et-al-2002}). Their result is 1.4~kpc, apparently
in better agreement with the extinction distance. Here we would
like to sound a word of caution: a point which was made already some
time ago by Steffen et al.~(\cite{Steffen-et-al-1997}) and again by
Sch\"{o}nberner~(\cite{Schoenberner-2001}). The outer rim of a PN is
defined by a shock front, the temporal displacement of which is not
given by a material velocity and is not easily derivable from the
Doppler splitting of the strongest PN emission lines. Hydrodynamic
modelling indicates that frequently the Doppler splitting is smaller
than the linear velocity of expansion in the plane of the sky. The
assumption that both are equal can easily lead to systematically too
small expansion distances, perhaps by a factor as large as 2. For that
reason we think that some more work is needed on the interpretation
of the angular expansion of PNs.

\subsection{Summary on distances}

The amount of information is too small to extract any solid
conclusion. The independent evidence would seem to provide support
to several spectroscopic distances, but there are a few discrepant
cases that need to be resolved.

Since we are primarily interested in testing the validity of the
theory of post-AGB evolution, a few more comments are relevant.
An interesting consequence of the extinction distances is that they
produce several central stars with extremely low luminosities, which
cannot be post-AGB stars (see Martin~\cite{Martin-1994}). Therefore
we may still have a severe problem with the classical interpretation
of several {\em other} CSPNs as post-AGB stars; similar to what we
found for NGC~2392.

In this situation we need more and better independent distance determinations,
good enough to convince everybody. For the moment, we find no compelling reason
to reject the spectroscopic distances, although we understand that some of them
are taken with skepticism. But we would expect the spectroscopic distances, if
based on an inadequate physical theory, to fail all together in a very systematic
way; not just a few of them wrong and all the others OK. And so we still expect
that the few conflicting cases may be resolved in favor of the spectroscopic
distances when more evidence is added.

\subsection{The constraints from the PN luminosity function}

There is another verification we can undertake, based on the behavior of
extragalactic PNs. They show a very characteristic luminosity function (PNLF),
with a well-defined limiting brightness, which has been successfully used for
extragalactic distance determinations (see, e.g., Jacoby and
Ciardullo~\cite{Jacoby-Ciardullo-1993}, Jacoby~\cite{Jacoby-1997}). We can try to
verify if our spectroscopic distances produce any overluminous PN; that would be
a nice argument supporting a smaller spectroscopic distance in that case.

Now one complication is that for extragalactic work the PNLF is
built using the normally very bright nebular emission [O\,{\sc iii}]
$\lambda$5007. Very low-excitation PNs do not contribute to the
bright end of the $\lambda$5007 PNLF. But it turns out that some
of our central stars belong to low-excitation PNs, implying that
the flux in $\lambda$5007 does not provide any useful limit. For
that reason we have decided to do the test using a recombination
line, namely H$\beta$. The problem is now that we do not have too
much observational information about the limiting H$\beta$ flux
in other galaxies: the only well-observed case is the LMC. But
we can try to supplement the observational LMC limit by a limit
obtained from numerical simulations of the PNLF: see M\'endez and
Soffner~(\cite{Mendez-Soffner-1997}). Their Figure~6 shows the observed
H$\beta$ LMC PNLF, compared with a simulated PNLF. Allowing for a
somewhat larger sample size in our Galaxy (see the effect of increasing
the sample size in Figure~10 of M\'endez and Soffner), we can estimate
that the brightest PNs in our Galaxy should have an absolute H$\beta$
magnitude of about $-2.3$ (the relation between observed H$\beta$
flux and H$\beta$ apparent magnitude is traditionally defined as
$m_\beta = -2.5 \log F_\beta - 13.74$).

The resulting absolute H$\beta$ magnitudes we derive using our
spectroscopic distances are listed in Table~\ref{tbl:distances}. There
is only one case at the limit: He~2-131, with $M_\beta = -2.4$. All the
other distances produce weaker absolute H$\beta$ magnitudes. Again
we find no strong reason to reject our spectroscopic distances,
although He~2-131 is admittedly at the very limit of acceptability.

\section{Masses of white dwarfs and pre-white dwarfs}
\label{sec:masses}

Probably the most severe conflict we have is the large number of very
massive CSPNs, in view of the known mass distribution of white dwarfs,
with a well-defined maximum at about 0.6~$\Msun$. Although this could
be used to argue against the credibility of our analysis, we would
like to point to the existence of some recent results involving very
massive pre-white dwarfs and white dwarfs. Most interesting is a report
by Kawaler~(\cite{Kawaler-2001}) who finds a wide range of pulsation
periods among H-deficient CSPNs, which he tentatively interprets
as due to a correspondingly large range of masses, from 0.52 to 1.2
$\Msun$. This looks surprisingly similar to our result, based on a
completely different observational technique applied to a completely
different sample of central stars (our stars have H-rich atmospheres).

Another study worth mentioning is by Napiwotzki et
al.~(\cite{Napiwotzki-et-al-1999}). They have determined masses for a
sample of 46 hot DA white dwarfs selected from the Extreme UV Explorer
(EUVE) and ROSAT Wide Field Camera bright source lists. They find
a peak mass of 0.59~$\Msun$, in agreement with many other studies,
but find a non-negligible fraction of white dwarfs with masses in
excess of 1~$\Msun$.

Yet another study by Silvestri et al.~(\cite{Silvestri-et-al-2001}),
dealing with a sample of 41 white dwarfs in wide binary systems,
finds a bimodal mass distribution with a second mass peak at 1.1
$\Msun$. They interpret this second peak, suspiciously close to twice
the mass of the first peak, as the result of binary mergers.

Therefore, our mass distribution, with its probable dependence on
very strong selection effects, is perhaps not as irreconcilable with
the rest of the evidence as we could have thought initially.

The conflict with the post-AGB evolutionary speeds is not too
important if we decide to accept a drastic departure from the relation
between luminosity and mass. In this case new stellar structure and
evolutionary calculations would be needed.

\section{Conclusions and perspectives}
\label{sec:conclusions}

We have applied our new model atmospheres, involving a much improved treatment of
blocking and blanketing by all metal lines in the entire sub- and supersonically
expanding atmosphere, to the analysis of a sample of 8 PN central stars. We have
shown how the new models can produce an essentially perfect fit to a multitude of
spectral features in the UV~spectra of the CSPNs. The fits lead us to determine a
set of stellar parameters including separate determinations of luminosity and
mass, allowing for the first time a full test of the post-AGB evolutionary
calculations. Surprisingly, we find drastic departures from the theoretical
post-AGB mass--luminosity relation.

The luminosities we derive for the stars of our sample lie in the
expected range, but we find a much larger spread in the masses, from
0.4 to 1.4~$\Msun$.  The resulting relation between CSPN mass
and luminosity deviates severely from that given by the theory of
post-AGB evolution.

For five out of the nine CSPNs of our sample we obtain masses near,
{\em but not above\/}, the critical Chandrasekhar mass limit for
white dwarfs.  Despite our sample most probably being influenced
by selection effects, this result nevertheless invites speculation
about the role of a group of CSPNs as precursors to the white dwarfs
believed to end up as Type~Ia supernovae.

We cannot at the moment offer a clear-cut explanation to the
discrepancy between radiation-driven wind theory confirmed by
UV-spectroscopy on the one hand and the theory of post-AGB stellar
evolution on the other (in particular the fact that from the former
we derive masses both larger and smaller than those predicted by
the latter); however, we point out a number of other independent
observational investigations (see section~\ref{sec:other-observations})
that have also found a similarly large spread (up to 1.2~$\Msun$)
in the CSPN/white-dwarf masses which cannot be explained by the
classical post-AGB evolutionary theory. Nevertheless, if we believe
both, the current evolutionary theory and the luminosities and masses
we have determined from the atmospheric models, then most of the
CSPNs of our sample have not followed a classical post-AGB evolution.

\begin{acknowledgements}
This work was supported by the
Sonder\-for\-schungs\-bereich~375 of the Deutsche Forschungsgemeinschaft
and by the DLR under grant 50~OR~9909~2.
\end{acknowledgements}

\end{document}

%% file: table1.tex
\small
\begin{tabular}{r@{~\,\,}lcccccr@{.}l@{\extracolsep{5pt}}r@{}l}
& & $\Teff$ & $R$ & & $M$ & $\logg$ &
\multicolumn{2}{c}{$\Mdot$} &
\multicolumn{1}{c}{$\vinf$} & \\
& \multicolumn{1}{c}{\raisebox{7pt}[-7pt]{Object}} & (K) & ($\Rsun$) &
\raisebox{7pt}[-7pt]{$\displaystyle \log {L \over \Lsun}$} &
($\Msun$) & (cgs) &
\multicolumn{2}{@{}c@{}}{($10^{-6}\Msun/{\rm yr}$)} &
\multicolumn{1}{@{}c@{}}{(km/s)} & \\[2pt] \hline
\multicolumn{11}{c}{our models} \\ \hline
& NGC 2392 & 40000 & 1.5 & 3.7 & 0.41 & 3.70 &       0&018 &  420 & \\
& NGC 3242 & 75000 & 0.3 & 3.5 & 0.53 & 5.15 &       0&004 & 2400 & \\
( & IC 4637 & 55000 & 0.8 & 3.7 & 0.87 & 4.57 &       0&019 & 1500 & ) \\
& IC 4593  & 40000 & 2.2 & 4.0 & 1.11 & 3.80 &       0&062 &  850 & \\
& He 2-108 & 39000 & 2.7 & 4.2 & 1.33 & 3.70 &       0&072 &  800 & \\
& IC 418   & 39000 & 2.7 & 4.2 & 1.33 & 3.70 &       0&072 &  800 & \\
& Tc 1     & 35000 & 3.0 & 4.1 & 1.37 & 3.62 &       0&021 &  900 & \\
& He 2-131 & 33000 & 5.5 & 4.5 & 1.39 & 3.10 &       0&35  &  450 & \\
& NGC 6826 & 44000 & 2.2 & 4.2 & 1.40 & 3.90 &       0&18  & 1200 & \\ \hline
\multicolumn{11}{c}{Kudritzki et al.~1997} \\ \hline
& NGC 2392 & 45000 & 2.5 & 4.4 & 0.91 & 3.6  & ~$\le$ 0&03  &  400 & \\
& NGC 3242 & 75000 & 0.6 & 4.0 & 0.66 & 4.7  & $\le$ 0&02  & 2300 & \\
& IC 4637  & 55000 & 1.3 & 4.1 & 0.78 & 4.1  & $\le$ 0&02  & 1500 & \\
& IC 4593  & 40000 & 2.2 & 4.0 & 0.70 & 3.6  &       0&1   &  900 & \\
& He 2-108 & 35000 & 3.2 & 4.1 & 0.75 & 3.3  &       0&24  &  700 & \\
& IC 418   & 37000 & 3.5 & 4.3 & 0.89 & 3.3  &       0&26  &  700 & \\
& Tc 1     & 33000 & 5.1 & 4.4 & 0.95 & 3.0  & $\le$ 0&1   &  900 & \\
& He 2-131 & 30000 & 5.5 & 4.3 & 0.88 & 2.9  &       0&9   &  500 & \\
& NGC 6826 & 50000 & 2.0 & 4.4 & 0.92 & 3.8  &       0&26  & 1200 & \\ \hline
\end{tabular}

%% file: table2.tex
\newdimen\digitwidth
\setbox0=\hbox{0}
\digitwidth=\wd0
\catcode`?=13
\def?{\kern\digitwidth}
\small
\begin{tabular}{lccrccccr}
& Mass & $\logg$ & \multicolumn{1}{c}{$F_\star$} & $V$ & $c$ &
distance & $-\log F({\rm H\beta})$ & \multicolumn{1}{c}{$M_\beta$} \\
\multicolumn{1}{c}{\raisebox{7pt}[-7pt]{Object}} & ($M_\odot$) & (cgs) & & & &
(kpc) & & \\ \hline
NGC 2392 & 0.41 & 3.70 &  7.48 & 10.53 & 0.16 & 1.67 & 10.29 &    0.5 \\ 
NGC 3242 & 0.53 & 5.15 & 14.69 & 12.10 & 0.09 & 1.10 & ?9.80 &    0.3 \\
IC 4637  & 0.87 & 4.57 & 10.64 & 12.47 & 1.10 & 1.01 & 11.24 &    1.6 \\
IC 4593  & 1.12 & 3.80 &  7.69 & 11.27 & 0.12 & 3.63 & 10.55 & $-$0.5 \\
He 2-108 & 1.33 & 3.70 &  7.49 & 12.82 & 0.40 & 6.76 & 11.41 & $-$0.4 \\
IC 418   & 1.33 & 3.70 &  7.49 & 10.00 & 0.32 & 2.00 & ?9.62 & $-$2.0 \\
Tc 1     & 1.37 & 3.62 &  6.38 & 11.38 & 0.36 & 3.73 & 10.66 & $-$0.8 \\
He 2-131 & 1.39 & 3.10 &  6.26 & 10.50 & 0.14 & 5.62 & 10.16 & $-$2.4 \\
NGC 6826 & 1.40 & 3.90 &  8.61 & 10.69 & 0.04 & 3.18 & ?9.97 & $-$1.4 \\
\end{tabular}